\documentclass{article}

\usepackage[a4paper]{geometry}
\usepackage{xcolor}
\usepackage[utf8]{inputenc} 
\usepackage{amsmath,amsopn,amsthm} 
\usepackage[mathscr]{eucal} 
\usepackage{float}
\usepackage{MnSymbol}
\usepackage{paralist}
\usepackage{array,tabularx}
\usepackage[hyperindex,pagebackref,final,unicode,naturalnames]{hyperref}
\hypersetup{
  colorlinks = true,
  linkcolor = red,
  anchorcolor = black,
  citecolor = blue, 
  filecolor = cyan,
  menucolor = red,
  runcolor = cyan,
  urlcolor = magenta,
  linkbordercolor = white,
  linktocpage = true
}

\makeatletter
\long\def\@makecaption#1#2{%
 \vskip\abovecaptionskip
 \sbox\@tempboxa{\small #1. #2}
 \ifdim \wd\@tempboxa >\hsize
		\footnotesize #1. #2
 \else
	 \global \@minipagefalse
	 \hb@xt@\hsize{\hfil\box\@tempboxa\hfil}%
 \fi
 \vskip\belowcaptionskip}
\makeatother

\renewcommand{\phi}{\varphi}

\usepackage{tikz}
\usetikzlibrary{positioning,calc,automata}

\tikzset{my loop/.style={->,to path={
.. controls +(70:0.8) and +(110:0.8) .. (\tikztotarget) \tikztonodes}},
my state/.style={circle,draw}}

\newtheorem{theorem}{Theorem}
\newtheorem{proposition}{Proposition}
\newtheorem{lemma}{Lemma}
\newtheorem{corollary}{Corollary}
\theoremstyle{definition}

\theoremstyle{remark}
\newtheorem{remark}{Remark}

\newtheoremstyle{note}
{0pt}
{0pt}
{}
{}
{\bfseries}
{.}
{.5em}
{}

\theoremstyle{note}
\newtheorem{ennote}[theorem]{Property}

\title{$\omega$-Automata}
\author{Thomas Wilke\\
  Kiel University\\
  email:\,\url{thomas.wilke@email.uni-kiel.de}}
\date{}

\begin{document}

\maketitle

\begin{abstract}
  This paper gives a concise introduction into the basic theory of $\omega$-automata (as of March 2014). The starting point are the different types of recurrence conditions, modes of operation (deterministic, nondeterministic, alternating automata), and directions (forward or backward automata). The main focus is on fundamental automata constructions, for instance, for boolean operations, determinization, disambiguation, and removing alternation. It also covers some algebraic aspects such as congruences for $\omega$-automata (and $\omega$-languages), basic structure theory (loops), and applications in mathematical logic.---This paper may eventually become a chapter in a handbook of automata theory.
\end{abstract}



\tableofcontents

\newcommand{\Extref}[1]{???}
\newcommand{\Occ}[1]{\text{occ}(#1)} 
\newcommand{\Aut}[1]{\mathscr{#1}} 
\newcommand{\LeftLeq}{\leq_{\text{lft}}} 
\newcommand{\LeftLt}{<_{\text{lft}}}

\section{Introduction}

The expression ``$\omega$-automata'' generally refers to automata that accept or reject $\omega$-words---infinite sequences of letters from some alphabet. They define $\omega$-languages---sets of $\omega$-words---just as ordinary automata define languages of finite words; they are means for working with $\omega$-languages, just as ordinary automata are means for working with languages of finite words.

The fundamental questions about $\omega$-automata are similar to the fundamental questions about automata on finite words. Can operations on languages, such as intersection, complementation, and projection, be performed on automata? What is the difference between nondeterminism and determinism? What is the descriptional complexity of various types of automata? Which types of automata can be transformed into other types and at which cost? Can automata be minimized efficiently? How can automata be compared? \dots\ 

Some of the techniques developed for finite words can be adopted in the infinite-word setting; some of the theory of $\omega$-automata is very similar to ordinary automata theory. There are, however, many interesting new aspects, which, most often, have something to do with what happens ``in the infinite''. As this interesting behavior ``in the infinite'' is already present in $\omega$-automata with a finite state space, this paper is limited to the theory of finite-state $\omega$-automata.

To convey the core ideas as crisp and clear as possible within the given space limits, some of the material is presented in an uncommon way, at the expense of continuity with prior work.

Excellent surveys that cover $\omega$-languages and $\omega$-automata in their
entire breadth, especially their relationship with mathematical logic, have been
written by Wolfgang Thomas, one in the late eighties \cite{Thomas1990}, and one
in the nineties \cite{thomas-languages-automata-logic-handbook-1997}. There is
also a comprehensive monograph by Dominique Perrin and Jean-\'{E}ric Pin
\cite{perrin-pin-infinite-words-2004}. This paper tries to be a concise introduction into the \emph{theory} of $\omega$-automata.

\subsection{\texorpdfstring{$\omega$}{ω}-Words}
\label{sec:omega-words}

$\omega$-Automata are devices that work on $\omega$-words rather than finite ones. Technically, an $\omega$-word \index{$\omega$-word} over an alphabet $A$ is a function $\omega \to A$, where $\omega$ stands for the set of natural numbers. In contrast, a finite word over $A$ is a function $[n] \to A$, where $n$ is a natural number and $[n]$ denotes the set $\{0, \dots, n-1\}$. When $u$ is a word, finite or infinite, then $\Occ u$ denotes the set of letters occurring in $u$; when $u$ is an $\omega$-word, then $\inf(u)$ denotes the set of letters occurring infinitely often in $u$.

There are essentially three concatenation operations involving $\omega$-words. First, given a finite word $u$ and an $\omega$-word $v$, the $\omega$-word obtained by appending $v$ to~$u$ is denoted $u \cdot v$---we speak of $\omega$-concatenation\index{$\omega$-concatenation}. Second, when $\langle u_0, u_1, \dots\rangle$ is an infinite sequence of finite nonempty words, then $u_0 \cdot u_1 \cdot \dots$ denotes the $\omega$-word obtained by concatenating all the $u_i$'s in the given order---we speak of $\omega$-product\index{$\omega$-product}. Finally, when $u$ is a finite nonempty word, then $u^\omega$ denotes $u \cdot u \cdot \dots$---we speak of $\omega$-power\index{$\omega$-power}. Note that it is legitimate to use the same symbol $\cdot$ for ordinary concatenation, $\omega$-concatenation, and $\omega$-product, because there are various laws of associativity that hold. As usual, the symbol ``$\cdot$'', representing the different forms of concatenation, is omitted in many contexts, and these operations are extended to sets of words in a straightforward fashion. An $\omega$-word $u$ is periodic if $u = v^\omega$ for some finite nonempty word $v$; it is ultimately periodic\index{ultimately periodic}\index{word!ultimately periodic} if $u = v w^\omega$ for finite words $v$ and $w$ with $w$ being nonempty. The convolution of $\omega$-words $u$ and $v$ over alphabets $A$ and $B$, respectively, is an $\omega$-word over the alphabet $A \times B$, denoted $u * v$ and defined by $(u * v)(i) = \langle u(i), v(i)\rangle$ for every $i$. 

One of the reasons why $\omega$-automata are applicable in various situations is that $\omega$-words can represent infinite objects and therefore $\omega$-automata can represent sets of infinite objects or even transform infinite objects into other infinite objects.  

When the alphabet $A$ is the binary alphabet, $[2]$, then an $\omega$-word can be identified with a subset of $\omega$, that is, with a set of natural numbers, more precisely, a word $u$ can be identified with $\{i \in \omega \mid u(i) = 1\}$. When the alphabet is $\bigtimes_{i<k} [2]$, the $k$-fold cartesian product of $[2]$, then an $\omega$-word can be identified with a $k$-tuple of sets of natural numbers.

Every $\omega$-word over $[2]$ represents a real number from the interval $[0,1]$ in a natural fashion, more precisely, $u \in [2]^\omega$ represents the number $\sum_i u(i) \,  2^{-i-1}$. Observe that some numbers are represented twice, for instance, $1/2$, which is represented by $1 0^\omega$ and by $01^\omega$. 

There are various ways to represent any real number by an $\omega$-word. One way is to consider only words $u$ over $[2]$ where $u(2i+2) = 0$ for almost all $i$ and then let $u$ represent $(-1)^{u(0)} (\sum_i u(2i+2) 2^i  + \sum_i u(2i+1) 2^{-i-1})$. So the letters at even positions determine the integer part, including the sign, and the letter at odd positions determine the fractional part. Another way is to use a larger alphabet, for instance, $[2] \times [2]$, and represent the integer part in one dimension and the fractional part in the other dimension.

In this paper, a binary tree is a prefix-closed subset of $[2]^*$; level $i$ of such a tree~$T$ is the set of vertices $T \cap [2]^i$; its width is $\sup_i |T \cap [2]^i|$. Given some $k$, the set of trees of width at most $k$ can be represented by $\omega$-words over a fixed alphabet. A simple such representation is the sequence of (representations of) its slices, where a slice is two consecutive levels together with their interconnections, that is, a slice looks like this: 
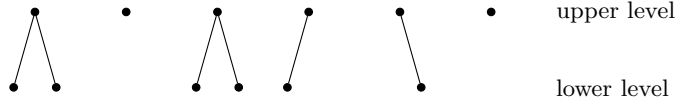
\begin{center}
 \small
 \begin{tikzpicture}[level distance=10mm, sibling distance=7mm, every node/.style={draw,circle,fill,inner sep=1pt},xscale=0.8]
	 \node at (-3,0) {} 
	 child {node {}}
	 child {node {}};
	 \node at (-1.5,0) {};
	 \node at (0,0) {} 
	 child {node {}}
	 child {node {}};
	 \node at (1.5,0) {} 
	 child {node {}}
	 child[missing];
	 \node at (3,0) {} 
	 child[missing]
	 child {node {}};
	 \node at (4.5,0) {}; 
	 \node [right,fill=white,draw=none] at (5.5,0) {upper level};
	 \node [right,fill=white,draw=none] at (5.5,-1) {lower level};
 \end{tikzpicture}
\end{center}
To represent this slice one could use the single ``letter'' $\langle 1,1\rangle\langle 0,0\rangle\langle 1,1\rangle\langle 1,0\rangle\langle 0,1\rangle\langle 0,0\rangle$. 

Not all $\omega$-words over the respective alphabet represent a tree, but every tree (of a given maximum width~$k$) can be represented. For instance, the tree denoted by $0^*(\epsilon + 1 + 11)$, which looks like a comb, is represented by $\langle 1,1\rangle \, (\langle 1,1 \rangle \langle 0,1 \rangle) \, (\langle 1,1 \rangle \langle 0,1 \rangle \langle 0,0 \rangle)^\omega$.

A labeled binary tree is a function $T \to A$ from a binary tree to an alphabet, and such trees, if restricted in their width, can also be represented by $\omega$-words, simply by augmenting the above representation by information about the labels. It is enough to encode in the representation of one slice the labels of the vertices in the ``upper'' level. 

What has just been said for infinite trees is also true for graphs to a certain extent. A leveled DAG \index{leveled DAG}\index{DAG!leveled} is a directed acyclic graph together with a partition of its vertex set into levels, more precisely, such a graph is given by a family $\{V^{(i)}\}_{i \in \omega}$ of pairwise disjoint vertex sets and an edge set $E \subseteq \bigcup_i V^{(i)} \times V^{(i+1)}$. The set of all vertices, $\bigcup_i V^{(i)}$, is denoted $V$; the elements of $V^{(i)}$ are the vertices on level $i$. Similarly to above, the width of such a DAG is $\sup_i |V^{(i)}|$. 

Given a natural number~$k$, leveled DAG's of width at most $k$ can be represented over a fixed alphabet, again by spelling out the sequence of its slices, where a slice represents a subgraph induced by two consecutive levels. Such a subgraph can, for instance, look like this:
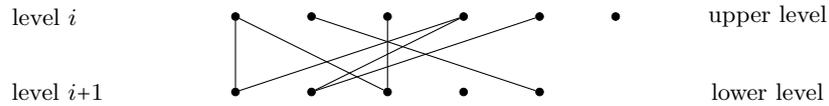
\begin{center}
 \small
 \begin{tikzpicture}[every node/.style={circle, draw, fill, inner sep=1pt}]
	 \node (a) at (0,0) {};
	 \node (b) at (1,0) {};
	 \node (c) at (2,0) {};
	 \node (d) at (3,0) {};
	 \node (e) at (4,0) {};
	 \node (f) at (5,0) {};

	 \node (A) at (0,-1) {};
	 \node (B) at (1,-1) {};
	 \node (C) at (2,-1) {};
	 \node (E) at (3,-1) {};
	 \node (D) at (4,-1) {};
	 \node[fill=white,draw=none] (F) at (5,-1) {};

	 \node [fill=white,draw=none,right of=f,xshift=10mm] {upper level};
	 \node [fill=white,draw=none,right of=F,xshift=10mm] {lower level};
	 \node [fill=white,draw=none,left of=a,xshift=-20mm,right] {level $i$};
	 \node [fill=white,draw=none,left of=A,xshift=-20mm,right] {level $i{+}1$};

	 \draw (a) -- (A) {};
	 \draw (a) -- (C) {};
	 \draw (c) -- (C) {};
	 \draw (b) -- (D) {}; 
	 \draw (d) -- (A) {};
	 \draw (d) -- (B) {};
	 \draw (e) -- (B) {};
 \end{tikzpicture}
\end{center}
This slice could be represented by the ``letter'' $\langle \{0, 2\}, \{4\}, \{2\}, \{0, 1\}, \{1\}, \{\}\rangle$, an enumeration of the adjacency sets of the vertices on the the upper level. We are only interested in leveled graphs up to isomorphism; so there are, in general, many representations for the same graph. Labeled leveled DAG's of bounded width can also be represented by $\omega$-words.

\section{Types of \texorpdfstring{$\omega$}{ω}-automata}

A finite-state nondeterministic $\omega$-automaton \index{finite-state $\omega$-automaton}\index{$\omega$-automaton!finite-state} over a given alphabet $A$ consists of
\begin{itemize}
\item a finite set of states, $Q$,
\item a set of initial states, $Q_I \subseteq Q$,
\item a set of transitions, $\Delta \subseteq Q \times A \times Q$, and
\item a representation of a recurrence condition.\index{recurrence condition}\index{condition!recurrence}
\end{itemize}
The last ingredient is the one that distinguishes an $\omega$-automaton from an ordinary finite-state automaton that works on finite words; it replaces the set of final states. This is necessary, because $\omega$-words have no end. But before recurrence conditions can be explained in detail, the notion of a run needs to be adapted to $\omega$-words. 

A run of an $\omega$-automaton on a given word $u \in A^\omega$ is an $\omega$-word $r \in Q^\omega$ such that $\langle r(i), u(i), r(i+1)\rangle \in \Delta$ holds for every $i \in \omega$. It is initial if $r(0) \in Q_I$ holds; it is accepting if it is initial and recurring, and what it means for $r$ to be recurring depends on the recurrence condition, as described in what follows. 

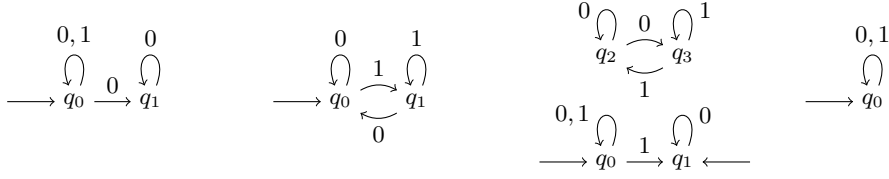
\begin{figure}
  \small
  \centering
 \begin{tikzpicture}
	 \node[] (q) {};
	 \node (q0) [right of=q] {$q_0$};
	 \node (q1) [right of=q0] {$q_1$}; 

	 \draw[->] (q) -- (q0);
	 \draw[->] (q0) to node[above] {$0$} (q1);

	 \draw[my loop] (q0) to node[above] {$0,1$} (q0);
	 \draw[my loop] (q1) to node[above] {$0$} (q1);

	 \node[] (p) [right of=q,xshift = 25mm] {};
	 \node (p0) [right of=p] {$q_0$};
	 \node (p1) [right of=p0] {$q_1$}; 

	 \draw[->] (p) -- (p0);
	 \draw[->,bend left] (p0) to node[above] {$1$} (p1);
	 \draw[->,bend left] (p1) to node[below] {$0$} (p0);

	 \draw[my loop] (p0) to node[above] {$0$} (p0);
	 \draw[my loop] (p1) to node[above] {$1$} (p1);

	 \node[] (r) [right of=p,xshift = 25mm,yshift=-8mm] {};
	 \node (r0) [right of=r] {$q_0$};
	 \node (r1) [right of=r0] {$q_1$}; 
	 \node (r2) [right of=r1] {}; 

	 \draw[->] (r) -- (r0);
	 \draw[->] (r0) to node[above] {$1$} (r1);
	 \draw[->] (r2) -- (r1);

	 \draw[my loop] (r0) to node[above, left=1mm] {$0,1$} (r0);
	 \draw[my loop] (r1) to node[above, right=1mm] {$0$} (r1);

	 \node[] (s) [above of=r,yshift = 4mm] {};
	 \node (s0) [right of=s] {$q_2$};
	 \node (s1) [right of=s0] {$q_3$}; 

	 \draw[->,bend left] (s0) to node[above] {$0$} (s1);
	 \draw[->,bend left] (s1) to node[below] {$1$} (s0);

	 \draw[my loop] (s0) to node[above, left=1mm] {$0$} (s0);
	 \draw[my loop] (s1) to node[above, right=1mm] {$1$} (s1);

	 \node[] (t) [right of=r,xshift = 25mm,yshift=8mm] {};
	 \node (t0) [right of=t] {$q_0$};

	 \draw[->] (t) -- (t0);

	 \draw[my loop] (t0) to node[above] {$0,1$} (t0);
 \end{tikzpicture}
 \caption{Different types of $\omega$-automata for the set of all $\omega$-words over $[2]$ with a finite number of occurrences of $1$, denoted $L_\text{fin1}$: the first automaton is forward nondeterministic and, if augmented with the B\"{u}chi condition $\{q_1\}$, recognizes $L_\text{fin1}$; the second one is forward deterministic and, if augmented with the co-B\"{u}chi condition $\{q_0\}$, the parity condition $\pi \colon q_0 \mapsto 2, q_1 \mapsto 1$, the Muller condition $\{\{q_0\}\}$, or a suitable Streett or Rabin condition, recognizes $L_\text{fin1}$; the third one is backward deterministic if used with the B\"{u}chi condition $\{q_1, q_3\}$ and then recognizes $L_\text{fin1}$; the fourth one is forward deterministic and recognizes $L_\text{fin1}$ if augmented with the transition-Muller condition $\{\{(q_0, 0, q_0)\}\}$.}
 \label{fig:simple-buechi}
\end{figure}

A simple type of recurrence condition is the B\"{u}chi type, which is represented as a set $B \subseteq Q$; a run $r$ is recurring if one of the states from $B$ occurs infinitely often in it, that is, if $\inf(r) \cap B \neq \emptyset$ holds. For instance, the language $L_\text{fin1}$ defined by $L_\text{fin1} = \{u \in [2]^\omega \mid \inf(u) = \{0\}\}$ (``$1$ occurs only finitely often'') is recognized by a B\"{u}chi automaton with two states, see Figure~\ref{fig:simple-buechi}. (As usual, an automaton recognizes the language consisting of the words the automaton has an accepting run for; when an automaton is denoted $\Aut A$, this language is denoted $\text L(\Aut A)$.)

Obviously, using a B\"{u}chi condition one can neither specify that two particular states occur infinitely often nor that a specific state occurs only finitely often. A type of recurrence condition which is sufficiently expressive in this respect is the Muller type. Such a condition is represented by a set $\mathscr M \subseteq \powerset(Q)$; a run $r$ is recurring if $\inf(r) \in \mathscr M$. In other words, one explicitly specifies which states occur infinitely often in a run and which are the ones that occur only finitely many times. The language $L_\text{fin1}$ can also be recognized by a Muller automaton with two states, see Figure~\ref{fig:simple-buechi}.

\newcommand{\wrt}{w.\,r.\,t.\xspace}

\index{Büchi recurrence condition}
\index{recurrence condition!Büchi}
\index{parity recurrence condition}
\index{recurrence condition!parity}
\index{Rabin recurrence condition}
\index{recurrence condition!Rabin}
\index{Streett recurrence condition}
\index{recurrence condition!Streett}
\index{Muller recurrence condition}
\index{recurrence condition!Muller}
\index{weak recurrence condition}
\index{recurrence condition!weak}
\index{co-Büchi recurrence condition}
\index{recurrence condition!co-Büchi}
\index{generalized Büchi recurrence condition}
\index{recurrence condition!generalized Büchi}
\index{transition recurrence condition}
\index{recurrence condition!transition}
There are essentially five different types of recurrence conditions that have been investigated traditionally, all explained in the upper part of Table~\ref{tab:acceptance-conditions} and named after their originators  \cite{buechi-decision-method-restricted-second-order-arithmetic-1962,rabin-decidability-of-second-order-theories-and-automata-on-infinite-trees-1969,Streett82,muller-infinite-sequences-and-finite-machines-1963} except for the parity condition \cite{mostowski-regular-expressions-for-infinite-trees-and-a-standard-form-of-automata-1984}. In the lower part of the table, there are three types of conditions derived from the B\"{u}chi condition \cite{muller-saoudi-schupp-alternating-automata-weak-monadic-theory-tree-its-complexity-1992,klarlund-progress-measures-complementation-omega-automata-applications-temporal-logic-1991,courcoubetis-vardi-wolper-yannakakis-memory-efficient-algorithms-verification-temporal-properties-1992}. 

The trivial recurrence condition, which is not mentioned in the table, considers every run recurrent. For instance, all representations of binary trees of a fixed width (over the alphabet indicated in Section~\ref{sec:omega-words}) are recognized by an automaton with trivial recurrence condition.

\begin{table}[!t]
 \renewcommand{\arraystretch}{1.2}
 \begin{tabularx}{\textwidth}{lll>{\raggedright\arraybackslash}X}
	 \hline
	 Name & & Format & Semantics\\ \hline 
	 B\"{u}chi & $B$ & $B \subseteq Q$ & $\inf(r) \cap B \neq \emptyset$\\
	 parity & $\mathscr P$ & $\pi \colon Q \to [2n]$, $n \in \omega$ & $\min(\inf(\pi \circ r)) \text{ mod } 2 = 0$\\
	 Rabin & $\mathscr R$ & $\mathscr R \subseteq \powerset(R) \times \powerset(R)$ & there exists $\langle L, U\rangle \in \mathscr R$ such that $\inf(r) \cap L = \emptyset$ and $\inf(r) \cap U \neq \emptyset$\\
	 Streett & $\mathscr S$ & $\mathscr S \subseteq \powerset(R) \times \powerset(R)$ & for every $\langle R, G\rangle \in \mathscr S$, if $\inf(r) \cap R \neq \emptyset$, then $\inf(r) \cap G \neq \emptyset$\\
	 Muller & $\mathscr M$ & $\mathscr M \subseteq \powerset(Q)$ & $\inf(r) \in \mathscr M$\\
	 \hline
	 weak & $W$ & $W \subseteq Q$, union of SCC's & $\inf(r) \subseteq W$\\
	 co-B\"{u}chi & $C$ & $C \subseteq Q$ & $\inf(r) \subseteq C$\\
	 gen.\ B\"{u}chi & $\mathscr G$ & $\mathscr G \subseteq \powerset(Q)$ & $B \cap \inf(r) \neq \emptyset$ for every $B \in \mathscr G$\\
	 \hline
 \end{tabularx}
 \caption{Recurrence conditions, ordered according their expressive power; ``gen.\ B\"{u}chi'' and ``SCC'' are abbreviations of ``generalized B\"{u}chi'' and ``strongly connected component'', respectively.}
 \label{tab:acceptance-conditions}
\end{table}

It is convenient to give names to the elements of the recurrence conditions: a state $q \in B$ is called a B\"{u}chi state, a pair $\langle L, U\rangle \in \mathscr R$ is called a Rabin pair, a pair $\langle R, G\rangle \in \mathscr S$  is called a Streett pair, a set $M \in \mathscr M$ is called a Muller set, a state $q \in W$ is called a weak state, a state $q \in C$ is called a co-B\"{u}chi state, and a set $B \in \mathscr G$ is called a B\"{u}chi set. The function $\pi$ is called priority function. Only weak, co-B\"{u}chi, and B\"{u}chi conditions have straightforward representations of size polynomial in the number of states of a given automaton.

Every type of recurrence condition is also considered in a transition variant, where states from~$Q$ are replaced by transitions from $\Delta$ and $\inf(r)$ is replaced by the set of triples $\langle q, a, q'\rangle$ for which there exist an infinite number of $i$ such that $\langle r(i), u(i), r(i+1)\rangle = \langle q, a, q'\rangle$. Transition variants come in more handy in certain situations, for instance, $L_\text{fin1}$ is recognized by a single-state transition-Muller automaton, see Figure~\ref{fig:simple-buechi}.\footnotemark[1]

Table~\ref{tab:conversions} shows how conditions of various types can be expressed in terms of conditions of other types: every B\"{u}chi condition may be viewed as a parity condition, which, in turn, can be viewed as a Rabin or Streett condition, and these can be viewed as Muller conditions. 

Unlike finite words, $\omega$-words are not symmetric (in the sense that there is no order isomorphism from the order of the natural numbers to its inverse). So when talking about determinism the direction makes a difference.%
\footnotetext[1]{In hindsight this paper should have been written using transition conditions throughout, in particular, Section~\ref{run trees} would profit much from this.}
\refstepcounter{footnote}
A forward deterministic automaton is one where $Q_I$ consists of exactly one state and $|\Delta(q, a)| = 1$ for all $q \in Q, a \in A$. (As usual, $\Delta(q,a)$ is used as an abbreviation of $\{q' \in Q \mid \langle q, a, q'\rangle \in \Delta\}$.) A backward deterministic automaton%
\footnote{In \cite{carton-michel-unambiguous-buechi-automata-2003}, where these automata were introduced, they are called ``complete unambiguous''; in \cite{perrin-pin-infinite-words-2004}, the attribute ``prophetic'' is used; in \cite{preugschat-wilke-effective-characterizations-of-simple-fragments-of-temporal-logic-using-carton-michel0automata-2013}, they are referred to as ``Carton-Michel automata''. The terminology used in this paper tries to be systematic.} 
\index{Carton-Michel automaton}
\index{prophetic automaton}
\index{automaton!Carton-Michel}
\index{automaton!prophetic}
is one where for every $\omega$-word over $A$, there is exactly one recurring run and $|\Delta(a, q')| = 1$ for all $a \in A, q' \in Q$. (Here, $\Delta(a, q')$ stands for $\{q \in Q \mid \langle q, a, q'\rangle \in \Delta\}$.) At times, when deterministic automata are used, the transition relation $\Delta$ is replaced by a transition function $\delta \colon Q \times A \to Q$ (forward automata) or $\delta \colon A \times Q \to Q$ (backward automata).

In general, a type of an $\omega$-automaton is given by a type of recurrence condition \emph{and} a type of mode, with the following modes being considered: nondeterministic (default), forward deterministic, backward deterministic, and alternating, defined in Section~\ref{sec:alternation}.

\begin{table}[b]
 \centering
 \renewcommand{\arraystretch}{1.2}
 \begin{tabularx}{\textwidth}{llX}
	 \hline
	 From & To & Conversion \\ \hline
	 B\"{u}chi $B$ & parity $\pi$ & $\pi(q) = 0$ for $q \in B$ and $\pi(q) = 1$ for $q \notin B$\\
	 parity $\pi$ & Rabin $\mathscr R$ & $\mathscr R = \{(\pi^{-1}(\{0, \dots, 2i-1\}), \pi^{-1}(\{0, \dots, 2i\})) \mid i + 1 < n\}$ \\
	 parity $\pi$ & Streett $\mathscr S$ & $\mathscr S = \{(\pi^{-1}(\{0, \dots, 2i+1\}), \pi^{-1}(\{0, \dots, 2i\})) \mid i + 1 < n\}$ \\
	 Rabin $\mathscr R$ & Muller $\mathscr M$ & $\mathscr M = $ set of all $Q' \subseteq Q$ such that there exists $\langle L, U\rangle \in \mathscr R$ satisfying $Q' \cap L = \emptyset$ and $Q' \cap U \neq \emptyset$ \\
	 Streett $\mathscr S$ & Muller $\mathscr M$ & $\mathscr M = $ set of all $Q' \subseteq Q$ such that, for every $\langle R, G\rangle \in \mathscr S$, if $Q' \cap R \neq \emptyset$, then $Q' \cap G \neq \emptyset$ \\ \hline
 \end{tabularx}
 \caption{Conversions between recurrence conditions}
 \label{tab:conversions}
\end{table}

The most fundamental result about $\omega$-automata compares the different types of $\omega$-automata with respect to their expressive power. As a yardstick, nondeterministic B\"{u}chi automata are used; the $\omega$-languages recognized by them are called regular $\omega$-languages, see Theorem~\ref{thm:regexps} for the origin of this terminology.

\begin{theorem}[equivalence of types of $\omega$-automata]
 \label{thm:fundamental}
 For every type of $\omega$-automaton, consider the class of $\omega$-languages recognized by automata of this type. Then all these classes coincide with the class of regular $\omega$-languages except for the classes corresponding to the following types: 
 \begin{inparaitem}[]
 \item forward deterministic generalized B\"{u}chi;
 \item forward deterministic B\"{u}chi; 
 \item forward deterministic, backward deterministic, and nondeterministic co-B\"{u}chi; 
 \item forward deterministic, backward deterministic, and nondeterministic weak.
 \end{inparaitem}
\end{theorem}

Much of $\omega$-automata theory revolves around Theorem~\ref{thm:fundamental}. The quest for good proofs of this theorem---efficient language-preserving transformations between automata of different types---has led to many interesting results. All types of $\omega$-automata are interesting in their own right; each one has its advantages and applications in specific contexts. 


\section{Basic properties of B\"{u}chi automata}

Some basic insights into $\omega$-automata can be derived from analyzing runs in a straightforward fashion, for instance, that regular languages can be defined by regular expressions of a certain type, that deterministic B\"{u}chi automata are less expressive than nondeterministic ones, and that complementation is problematic for B\"{u}chi automata.

\subsection{\texorpdfstring{$\omega$}{ω}-Regular expressions}

An $\omega$-regular expression \index{$\omega$-regular expression}\index{expression!$\omega$-regular} \cite{muller-infinite-sequences-and-finite-machines-1963} is of the form 
\begin{align}
 r_0 \cdot s_0^\omega + \dots + r_{n-1} \cdot s_{n-1}^\omega \enspace,
\end{align}
with $n$ being a natural number and the $r_i$'s and the $s_i$'s being ordinary regular expressions. The semantics is the obvious one.

Since the empty set can be denoted by an empty expression ($n = 0$) and since our definition of $\omega$-power is only defined for (sets of) nonempty finite words (see Section~\ref{sec:omega-words}), it is reasonable to require that the $s_i$'s be  built from the letters of the alphabet, ``$+$'' (for union), ``$\cdot$'' (for concatenation), and ``${}^+$'' (for finite positive iteration). It is also reasonable to allow that individual $r_i$'s are omitted.

\begin{theorem}[$\omega$-regular expression \cite{buechi-decision-method-restricted-second-order-arithmetic-1962}]
 \label{thm:regexps}
 Every $\omega$-language recognized by a B\"{u}chi automaton is denoted by an $\omega$-regular expression and vice versa.
\end{theorem}

For the proof, assume a B\"{u}chi automaton is given. The insight needed is that for every accepting run $r$ there are some state $q \in B$ and an infinite sequence $\langle i_0, i_1, \dots\rangle$ of positions such that $r(0)$ is initial, $i_0 < i_1 < \dots$, and $r(i_j) = q$ for every $j$. This motivates the following definition. For states $q, q'$, let $L_{q,q'}$ be the language recognized by the ordinary automaton on finite words with $q$ as initial and $q'$ as final state. Then the language recognized by the B\"{u}chi automaton is 
\begin{align} 
 \bigcup_{q \in Q_I, q' \in B} L_{q,q'} (L_{q',q'} \setminus \{\epsilon\})^\omega\enspace.
\end{align}
This representation can be turned into an $\omega$-regular expression using techniques known from finite-state automata on finite words.

\newcommand{\New}[1]{q_\text{new}}

For the proof of the converse, first observe that it is enough to show that an expression of the form $r \cdot s^\omega$ (meaning $n = 1$) denotes a language recognized by a B\"{u}chi automaton, because the class of languages recognized by B\"{u}chi automata is closed under union: the disjoint union of two given B\"{u}chi automata is a B\"{u}chi automaton recognizing the union of the languages recognized by the given automata. So assume $\mathscr A$ and $\mathscr B$ are finite-state automata on finite words recognizing the languages denoted by $r$ and $s$, respectively. A B\"{u}chi automaton for $r \cdot s^\omega$ is obtained by modifying the disjoint union of $\mathscr A$ and $\mathscr B$ as follows. First, an additional state $\New q$ is added. Second, for every transition $\langle q, a, q'\rangle$ in~$\mathscr A$ where $q'$ is final, the transition $\langle q, a, \New q\rangle$ is added. Third, for every transition $\langle q, a, q'\rangle$ in~$\mathscr B$ where $q$ is initial, the transition $\langle \New q, a, q'\rangle$ is introduced. Fourth, for every transition $\langle q, a, q'\rangle$ in $\mathscr B$ where $q'$ is final, the transition $\langle q, a, \New q\rangle$ is added. Finally, every final state looses its status as final state; every initial state of $\mathscr B$ looses its status as initial state; the state $\New q$ becomes the only B\"{u}chi state; if one of the initial states of $\Aut A$ was final (the empty word was accepted), then $\New q$ becomes initial, too.\qed

From the above proof, it immediately follows:

\begin{remark}
 \label{thm:basic}
 \begin{inparaenum}
 \item Every nonempty regular $\omega$-language contains an ultimately
	 periodic word.
 \item The emptiness problem for B\"{u}chi automata is decidable nondeterministically in logarithmic space, by a simple graph search.
 \end{inparaenum}
\end{remark}

\subsection{Co-B\"{u}chi and deterministic B\"{u}chi automata}
\label{sec:deterministic-buchi}

Dis- and reassembling runs is a simple but powerful technique in the context of $\omega$-automata, which can, for instance, be used to show that co-B\"{u}chi automata and deterministic B\"{u}chi automata are weaker than nondeterministic ones:

\begin{proposition}
 \label{thm:co-buchi}
 \begin{inparaenum}
 \item The language denoted by $((0+1)^* 0)^\omega$ cannot be recognized by a co-B\"{u}chi automaton.
 \item The language denoted by $(0+1)^* 1^\omega$, which is the complement of the language denoted by $((0+1)^* 0)^\omega$, cannot be recognized by a deterministic B\"{u}chi automaton. 
 \end{inparaenum}
\end{proposition}

For the proof of the first part, assume a co-B\"{u}chi automaton with $n$ states recognizes the language. Then it accepts the word $(1^n0)^\omega$, say $r$ is an accepting run. There is some~$k$ such that all letters in the segment $r[(n+1)k,(n+1)(k+1))$ are co-B\"{u}chi states. (As usual, if $u$ denotes an $\omega$-word, then $u[i,j)$ denotes $u(i)u(i+1) \dots u(j-1)$.) Because this segment has $n+1$ positions, there are $i$ and $j$ such that $i<j \leq n$ and $r((n+1)k+i) = r((n+1)k+j)$, which means $r[0,(n+1)k+i)r[(n+1)k+i,(n+1)k+j)^\omega$ is an accepting run of the automaton on $(1^n0)^k1^\omega$---a contradiction.

For the proof of the second part, assume a deterministic B\"{u}chi automaton recognizes the language denoted by $(0+1)^* 1^\omega$. By complementing its B\"{u}chi set and viewing it as a co-B\"{u}chi set, one obtains a co-B\"{u}chi automaton for the language denoted by $((0+1)^* 0)^\omega$---a contradiction to the first part.\qed

Proposition~\ref{thm:co-buchi} shows that the complementation procedure known from finite-state automata (first determinize, then negate the ``final condition''---does not work for B\"{u}chi automata, because the following two transformations are not possible in general: 
\begin{inparaitem}[]
\item from a nondeterministic B\"{u}chi automaton to an equivalent deterministic B\"{u}chi automaton;
\item from a deterministic B\"{u}chi automaton to a deterministic B\"{u}chi automaton for the complement of the language recognized.
\end{inparaitem}
There are fundamental differences between $\omega$-automata and ordinary ones. 

As complementation and determinization are important operations on automata in general, much of the work on $\omega$-automata deals with them and so does this paper. B\"{u}chi was the first to show that nondeterministic B\"{u}chi automata are closed under complementation \cite{buechi-decision-method-restricted-second-order-arithmetic-1962}; Safra's construction \cite{safra-complexity-omega-automata-1988}\index{Safra's construction} was the first with a worst-case state complexity of $\theta(n)^n$, which is optimal \cite{Michel88}. The first determinization construction for B\"{u}chi automata, transforming a nondeterministic B\"{u}chi automaton into an equivalent forward deterministic Rabin automaton, was given by McNaughton~\cite{mcnaughton-testing-and-generating-infinite-sequences-1966}; again, Safra's construction was the first with a worst-case state complexity of $\theta(n)^n$, which is, again, optimal \cite{loeding-optimal-bounds-for-transformations-omega-automata-1999}. 
The development with regard to complementation up to the year 2007 is described very nicely in~\cite{vardi-buechi-complementation-saga-2007}.

\section{Basic constructions}

In this section of introductory technical nature, some important basic constructions are described. They exhibit parallels to the situation with finite words, but demonstrate also distinctive features of $\omega$-automata. 

\subsection{Products of B\"{u}chi automata}
\label{sec:products}

A simple operation known from nondeterministic finite-state automata is the disjoint union of two automata, which yields a non-deterministic automaton recognizing the union of the two languages recognized by the two given automata. This works for $\omega$-automata exactly in the same way, provided the two automata have recurrence conditions of the same type.

Another simple operation known from finite-state automata is the product of two automata: it can be used to construct an automaton recognizing the intersection or the union of the two languages recognized by the given automata. The adaptation to $\omega$-words is possible, but not straightforward. In particular, if the recurrence condition is a B\"{u}chi condition, the problem arises that B\"{u}chi states may not be visited simultaneously, which, in the worst case, may result in an automaton not accepting a single word, while the intersection of the two languages may be the set of all $\omega$-words over the given alphabet.

The problem can be overcome by adding one bit to the state space. More precisely, a state in the adjusted product is of the form $\langle q_0, q_1, b\rangle$, where $b \in [2]$. The transition relation is chosen in such a way that $b = 1$ if there was a prior position with a B\"{u}chi state in the second component, but no position in between with a B\"{u}chi state in the first component. The word is accepted if a state $\langle q_0, q_1, 1\rangle$ with $q_0 \in B_0$ occurs infinitely often, that is, $B_0 \times Q_1 \times \{1\}$ is the B\"{u}chi set.

More precisely, assume there is a transition $\langle q_0, a, q_0'\rangle$ in the first automaton and a transition $\langle q_1, a, q_1'\rangle$ in the second one. This gives rise to a transition from $\langle q_0, q_1, b\rangle$ to $\langle q_0', q_1', b'\rangle$, where $b'$ is defined by: if $q_1 \in B_1$, then $b' = 1$; if $b = 1$ and $q_0 \in B_0$ and $q_1 \notin B_1$, then $b' = 0$; in all other cases, $b' = b$.

The constructed automaton is forward deterministic, provided the given automata are forward deterministic. When the transitions are reversed and the given automata are backward deterministic, then it is backward deterministic \cite{carton-michel-unambiguous-buechi-automata-2003}. The construction can also be used to turn a generalized B\"{u}chi automaton into an ordinary one, resulting in an automaton with $kn$ states, assuming the given automaton has $n$ states and $k$ B\"{u}chi sets.

\subsection{Automata with output and cascades}

In various situations, it is very helpful to consider $\omega$-automata with output\index{$\omega$-automaton!with output}, which have an extra output function $\lambda \colon \Delta \to O$, where $O$ is some alphabet. The relation defined by such an automaton is the relation between $A^\omega$ and $O^\omega$ which contains a pair $\langle u, v\rangle$ if there is an accepting run $r$ of the automaton on $u$ such that $v(i) = \lambda(\langle r(i), u(i), r(i+1)\rangle)$ for every~$i$.

A simple example is the relation which holds between a binary tree and an $\omega$-word if, and only if, the prefixes of the $\omega$-word form an infinite rooted path in the tree.  When the trees considered are of width at most $k$, then the states of a suitable automaton can be chosen to be elements of $[k]$, representing the vertex on the current level that is chosen to be part of the path. There is a transition $\langle i, a, j\rangle$ if vertex $j$ is a successor of vertex $i$ in slice $a$, and $\lambda(\langle i, a, j\rangle) = 0$ if $j$ is a left successor and else $\lambda(\langle i, a, j\rangle) = 1$. 

Just as in the theory of finite-state automata on finite words, an $\omega$-automaton with output can be composed with an $\omega$-automaton (with or without output) by using the output of the first automaton as input for the second automaton---one speaks of a cascade\index{cascade!of$\omega$-automata} or of cascading. For instance, if the above automaton is cascaded with a B\"{u}chi automaton recognizing $(1^*0)^\omega$, then the resulting automaton recognizes the trees that have a rooted path with infinitely many left successors. 

When cascading two finite-state automata on finite words a state of the resulting automaton is a pair consisting of a state of the first automaton and a state of the second one; when cascading $\omega$-automata one needs to be careful about the recurrence condition. For instance, if two B\"{u}chi automata are cascaded, then the result can be chosen to be a generalized B\"{u}chi automaton or a B\"{u}chi automaton with a third component, consisting of a single bit, for combining the two B\"{u}chi recurrence conditions as described in~Section~\ref{sec:products}.

\subsection{The breakpoint construction}
\label{sec:breakpoint}
\index{breakpoint construction}

A more important example for a B\"{u}chi automaton with output is an automaton which defines the function that maps each leveled DAG to the subgraph which is composed of the finitary vertices of the DAG, or, dually, the infinitary vertices. A vertex is called finitary\index{finitary!vertex of a DAG}\index{infinitary!vertex of a DAG} if it has only a finite number of descendants, else it is called infinitary. For DAG's of finite width, which we only consider, being infinitary is equivalent to being on an infinite path.

A B\"{u}chi automaton can guess which vertices on a level of a given DAG are finitary and which are not, and it can check that vertices guessed infinitary are indeed infinitary ones by forcing, via the transition relation, each infinitary vertex to have a successor. The problem is to verify that every vertex guessed finitary is indeed finitary. All successors of a vertex guessed finitary must be guessed finitary and this can be enforced by the transition relation, but this condition is only necessary and not sufficient. 

To solve the problem a construction referred to as breakpoint construction can be used. A ``breakpoint automaton'' works in phases. When a phase starts, all vertices on the current level guessed to be finitary are stored in some set, the verification set. During a phase, the verification set is updated from level to level by replacing the vertices in it by their successors. If the guesses were correct, the verification set becomes empty at some point and the phase ends successfully. During a phase, all vertices newly guessed finitary are stored in some other set---they are put on hold. When a new phase starts, the vertices put on hold previously are moved into the verification set. For all guesses to be correct, every phase has to end successfully.

A state of a breakpoint automaton is of the form $\langle V, H\rangle$, where $V$ is the current verification set and $H$ is the set of states currently put on hold. A transition is of the form $\langle\langle V, H\rangle, a, \langle V', H'\rangle\rangle$ and must satisfy the following conditions, which are all phrased with respect to the slice $a$ being read:
\begin{compactitem}
\item The sets $V$ and $H$ are disjoint sets of vertices of the upper level.
\item The sets $V'$ and $H'$ are disjoint sets of vertices of the lower level.
\item Every vertex on the upper level not in $V \cup H$ has a successor on the lower level not in $V' \cup H'$.
\item If $V \neq \emptyset$, then $V'$ is the set of all successors of the vertices in $V$, and $H'$ contains at least all the successors of the vertices in $H$ which do not belong to $V'$.
\item If $V = \emptyset$, then $V'$ is the set of all successors of the vertices in $H$.
\end{compactitem}
A state is initial if $H = \emptyset$; it is a B\"{u}chi state if $V = \emptyset$. 

To a transition as above, the output function assigns the slice which is obtained from~$a$ by restricting it to the vertices from $V \cup V' \cup H \cup H'$ or, dually, to the other vertices.

\begin{theorem}[breakpoint construction \cite{miyano-hayashi-alternating-finite-automata-on-omega-words-2-1984}]
 \label{thm:breakpoint}
 For every $k$, the breakpoint construction yields a B\"{u}chi automaton with $3^k$ states outputting, for every leveled DAG of width at most $k$, the subgraph of its finitary [infinitary]  vertices.
\end{theorem}

The breakpoint construction is used in Sections~\ref{sec:complementation}, \ref{sec:disambiguation}, \ref{thm:weak-characterization}, and \ref{sec:weak-alternation}.

\subsection{The lift construction}
\index{lift construction}

For backward deterministic automata, the previous task---computing the finitary or infinitary vertices in a leveled DAG---can be solved using a construction here referred to as lift construction. If one knows the finitary vertices of a leveled DAG on one level, one can determine the finitary vertices on the previous level in a deterministic fashion: a vertex on the previous level is finitary if, and only if, all its successors are finitary (in particular, a vertex without successors is finitary). The naive approach for constructing a backward deterministic automaton for determining the finitary vertices is to use states which have one bit for every vertex on the current level, indicating whether the vertex is finitary or not, and to use the above rule as a backward transition function. The problem is that this construction may overapproximate, because when in a run on a graph with only infinitary vertices all vertices are assumed to be finitary the above rule is obeyed.


To overcome the problem it is important to realize that for every finitary vertex $v$, say on level $l$, there is some smallest level $i > l$ without a descendant of $v$. We call this level the extinction level of $v$, denote it by $\text{el}(v)$, and use it to rank~$v$. 

\begin{figure}[t]
 \small
 \hfill
 \protect\begin{minipage}[t]{0.5\linewidth}
 \begin{tikzpicture}[baseline=(current bounding box.north),
	 node distance=5mm, every node/.style={circle,inner sep=1pt}]
	 \node (a0) {0};
	 \node[xshift=5mm] (b0) [right of=a0] {0};
	 \node[xshift=5mm] (c0) [right of=b0] {\bf 1};
	 \node[xshift=5mm] (d0) [right of=c0] {0};
	 \node[xshift=5mm] (e0) [right of=d0] {0};
	 \node[xshift=5mm] (f0) [right of=e0] {};

	 \node[xshift=-2mm,yshift=3mm] (aa) [below of=a0] {};
	 \node[xshift=2mm,yshift=3mm] (fa) [below of=f0] {};
	 \draw (aa) -- (fa);


	 \node (ad) [below of=a0] {};
	 \node (bd) [below of=b0] {};
	 \node (cd) [below of=c0] {2};
	 \node (dd) [below of=d0] {1};
	 \node (ed) [below of=e0] {0};
	 \node (fd) [below of=f0] {};

	 \node (a1) [above of=a0] {2};
	 \node (a2) [above of=a1] {0};

	 \node (b1) [above of=b0] {0};
	 \node (b2) [above of=b1] {0};
	 \node (b3) [above of=b2] {2};

	 \node (c1) [above of=c0] {0};
	 \node (c2) [above of=c1] {\bf 1};
	 \node (c3) [above of=c2] {\bf 2};

	 \node (d1) [above of=d0] {\bf 1};
	 \node (d2) [above of=d1] {0};

	 \node (e1) [above of=e0] {?};
	 \node (e2) [above of=e1] {?};
	 \node (e3) [above of=e2] {0};

	 \node (f1) [above of=f0] {};
	 \node (f2) [above of=f1] {};
	 \node (f3) [above of=f2] {};

	 \node (g1) [right of=f1] {\dots};


	 \draw[->] (a0) -- (b1);
	 \draw[->] (a1) -- (b3);
	 \draw[->] (a2) -- (b2);

	 \draw[->] (b0) -- (c1);
	 \draw[->] (b1) -- (c1);
	 \draw[->] (b2) -- (c1);
	 \draw[->] (b3) -- (c3);

	 \draw[->] (c0) -- (d2);
	 \draw[->] (c2) -- (d2);
	 \draw[->] (c2) -- (d0);
	 \draw[->] (c3) -- (d1);
	 \draw[->] (c3) -- (d2);

	 \draw[->] (d1) -- (e0);

	 \draw[->] (e1) -- (f1);
	 \draw[->] (e2) -- (f2);
	 \draw[->] (e1) -- (f3);
 \end{tikzpicture}  
 \end{minipage}
 \hfill
 \protect\begin{minipage}[t]{0.4\linewidth}
	 \protect\caption{Beginning of a leveled DAG with vertices labeled by their extinction ranks and critical values at the bottom.  Lifts are in bold; question marks indicate vertices with ranks which cannot be determined from the visible part of the DAG.}
	 \label{fig:ext-ranks}
	 \vspace*{\fill}
 \end{minipage}
 \hspace*{\fill}
\end{figure}
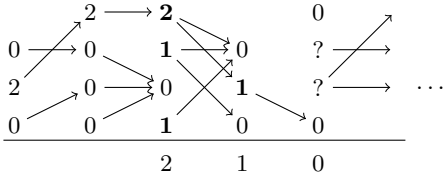

To this end, assume a vertex $v$ is on some level $l$. The rank of $v$ measures how difficult it is to get from $v$ to its extinction level, more concretely, how ``wide'' the part of the run DAG is which one needs to pass by while moving from $v$ to its extinction level. For every level $i$ between $l$ and the extinction level of $v$, that is, for every $i$ with $l \leq i < \text{el}(v)$, let $W_i$ contain all extinction levels of vertices on level~$i$, but only the ones which are before the extinction level of~$v$, that is, $W_i = \{\text{el}(w) \mid w \in V^{(i)} \text{ and } \text{el}(w) < \text{el}(v)\}$. The maximum of the cardinalities of the $W_i$'s is the extinction rank of~$v$ and denoted $\text{er}(v)$. Formally, $\text{er}(v) = \max\{|W_i| \mid l \leq i < el(v)\}$. For an illustration, see Figure~\ref{fig:ext-ranks}.

\begin{remark}
 \label{thm:ext-ranks}
 \begin{inparaenum}
 \item The extinction rank of a vertex in a leveled graph of width at most $k$ is at most $k-1$.
 \item On every maximum path starting in a finitary vertex the extinction rank is monotone descending and eventually reaches~$0$.
 \end{inparaenum}
\end{remark}

In the lift construction, the backward deterministic automaton determines, for each vertex, its extinction rank, that is, a state of the automaton maps each vertex on the current level to a value in $[k] \cup \{\infty\}$, where $\infty$ is used for infinitary vertices. It is possible to define a backward deterministic transition function accordingly, because the extinction ranks of the vertices on one level can be determined from the structure of the respective slice and the extinction ranks of the vertices on the next level, as described in what follows. 

Let $U$ be the set of vertices on the upper level of a slice~$a$. For every $v \in U$, let $m_v$ be the maximum of all values $\text{er}(v')$ where $v'$ is a successor of $v$ in the slice $a$; by convention, $m_v = -1$ if $v$ has no successor. If there is no $v \in U$ with $m_v = -1$, then $\text{er}(v) = m_v$ for all $v \in U$. If there is such a $v$, then the largest number $c$ such that $[c] \subseteq \{m_v \mid v \in U\}$ is called the critical value of the upper level. For every $v \in U$ with $m_v \geq c$, the equation $\text{er}(v) = m_v$ still holds. For every other $v \in U$, the values are ``lifted'': $\text{er}(v) = m_v + 1$.

If extinction ranks are used, then an overapproximation as described above can be avoided by adding an appropriate recurrence condition, more precisely, a generalized transition-B\"{u}chi condition. For every rank $i$, there is a transition-B\"{u}chi set $B_i$ which includes all transitions in which $i$ does not occur as a value on the upper level or $i$ is less than the critical value (and thus lifted), see Remark~\ref{thm:ext-ranks}(2) and Figure~\ref{fig:ext-ranks}. 

\begin{theorem}[lift construction   
 \cite{carton-michel-unambiguous-buechi-automata-2003}]
 \label{thm:shift}
 The lift construction yields, for every $k$, a backward deterministic generalized transition-B\"{u}chi automaton with at most $(k+1)^n$ states outputting, for every leveled DAG of width at most $k$, the subgraph of its finitary [infinitary] vertices.  
\end{theorem}

Note that the above approach is very versatile. If, for instance, one wants to determine the vertices which have at least one descendant with no successors, which one could call weakly finitary vertices, then one can take the same approach, replacing maximization by minimization. 

The above description of the lift construction is somewhat technical because of the measure introduced; a more ``automatic'' description follows. A state is a sequence $P_0 \dots P_{m-1}$ of nonempty pairwise disjoint sets of vertices. Assume a letter $a$ (a slice) is read backwards. Then the new state is determined in two steps. First, the sequence $P'_{-1} P'_0 \dots P'_{m-1}$ is determined where 
\begin{inparaenum}[(i)]
\item $P_{-1}'$ consists of all vertices $v$ on the upper level of $a$ without
  successors and 
\item $P_i'$, for $i \geq 0$, consists of all such vertices with
  some successor in $P_i$, but no successor in $P_{i'}$ for any $i' < i$.
\end{inparaenum}
Second, the new state is obtained from $P'_{-1} P'_0 \dots P'_{m-1}$ by removing all empty entries. The recurrence condition is, again, a generalized transition-B\"{u}chi condition: for every~$i$ there are infinitely many transitions with $i>m$ or $P_{i'}' \neq \emptyset$ for $i' < i$. The vertices that occur in the states are exactly the finitary ones.

\enlargethispage{\baselineskip}

The lift construction is used for different purposes in Section~\ref{sec:backward-determinization}.

\subsection{Latest appearance records}
\label{sec:lar}
\index{latest appearance record}\index{LAR!latest appearance record}\index{latest appearance automaton}\index{LAA!latest appearance automaton}

Given an alphabet $A$ and a special symbol $\$$ not in $A$, the latest appearance automaton (LAA) is a forward deterministic automaton with states being words over $A \cup \{\$\}$ where every letter from $A$ occurs at most once and $\$$ occurs exactly once. One such word is called a latest appearance record (LAR) and the part to the right of ``\$'' is its frame. 

The initial state of the LAA is the one-letter word $\$$; the recurrence condition is trivial; the transition function $\delta$ is defined as follows.  When $u$ is a state of the form $v\$v'$ and $a$ is a letter of the alphabet occurring in $vv'$, say $vv' = waw'$, then $\delta(u,a) = w\$w'a$. When~$a$ does not occur in $vv'$, then $\delta(u,a) = vv'\$a$. So the order in which the letters occur in the current state of the automaton is the order of their latest appearances in the prefix of the given word read so far, with all letters in the frame of the current state being the ones that have occurred since the previous occurrence of the letter just read. From this, the following can be derived.

\begin{remark}
 \cite{buchi-landweber-1969-a,gurevich-harrington-1982}
 Consider the frames of maximal length among all frames occurring infinitely often in the run of the LAA on a given word. Then all theses frames contain the same letters and these are exactly the ones occurring infinitely often in the given word.
\end{remark}

An interesting application of the latest appearance record is the transformation of a given Muller automaton into an equivalent parity automaton. First, the Muller condition is removed (and replaced by the trivial recurrence condition). Second, the automaton is augmented by the trivial output function, which simply outputs the current state. Third, the generated automaton is cascaded with the LAA over the state set of the Muller automaton. Finally, assuming the automaton has $n$ states, a priority function is added that assigns each state $\langle q, v\$v'\rangle$ the priority $2n - 2|v'|$ if $\Occ{v'}$ is a Muller set and else $2n - 2|v'| + 1$.

\begin{theorem}[Muller to parity]
 \label{thm:muller-to-parity}
 \cite{mostowski-regular-expressions-for-infinite-trees-and-a-standard-form-of-automata-1984}
 For every forward deterministic [non-deterministic] Muller automaton with $n$ states there is an equivalent forward deterministic [non-deter\-min\-is\-tic] parity automaton with $(n+1)!$ states and at most $2n$ priorities.
\end{theorem}

A refined construction, saving priorities if possible, is presented in Section~\ref{sec:parity-index}.

\section{Run DAG's of B\"{u}chi automata}
\label{sec:dags}

B\"{u}chi automata, in general, are nondeterministic automata, in other words, there may be several runs of a given B\"{u}chi automaton on a given word. These runs have to be considered at the same time if, for instance, one wants to turn a B\"{u}chi automaton into a B\"{u}chi automaton for the complement of the language recognized, because not to accept means \emph{all} initial runs are not recurrent. 

There are essentially two global structures that have been investigated for arranging all runs of a B\"{u}chi automaton in a concise way: DAG's and trees. The former are treated in this section, the latter in the next one. Applications are complementation, determinization, and disambiguation (defined in Section~\ref{sec:disambiguation}). 

Assume a B\"{u}chi automaton is given. The run DAG \index{run DAG}\index{DAG!run} of a given $\omega$-word $u$ is the leveled graph with levels $\{Q \times \{i\}\}_{i \in \omega}$ and edges $\langle \langle q, i \rangle, \langle q', i+1\rangle \rangle$ for $\langle q, u(i), q' \rangle \in \Delta$. Its width is the number of states of the given automaton. 

Often, it is useful to think of a run DAG as a graph labeled with elements from $Q$; in this section, it is sufficient to think of it as providing only information about whether the state component of a vertex is an initial or a B\"{u}chi state. Technically, the DAG is labeled with elements from $\powerset(\{I, B\})$ and we say it is $\{I,B\}$-tagged; if a vertex is labeled with a letter $a$ and $I \in a$, we say it is $I$-tagged, and, analogously, if $B \in a$, we say it is $B$-tagged.

A vertex of an $\{I,B\}$-tagged DAG is called $B$-recurring if a path with an infinite number of $B$-tagged vertices starts in it; it is called $B$-free if none of its descendants (including itself) is $B$-tagged. The ultimate width of such a DAG is the limes inferior of the number of non-$B$-recurring infinitary vertices on a given level.

\begin{remark}
	An $\omega$-word is accepted by a B\"{u}chi automaton if, and only if, there is an $I$-tagged $B$-recurring vertex on level~$0$ of the run DAG of the word.
\end{remark}

The main insight needed about $\{I,B\}$-tagged DAG's (or simply $\{B\}$-tagged DAG's) of finite width is that they can be decomposed in a simple manner. Consider the following operation, here called peeling. First, remove all finitary vertices; second, remove all $B$-free vertices. Peeling does not remove any $B$-recurring vertex, and if it does not change the DAG at all, then all vertices are $B$-recurring, because every vertex has a strict $B$-tagged descendant. 
Moreover, if there are non-$B$-recurring infinitary vertices, then peeling decreases the ultimate width by at least one, as explained in what follows.

Consider a non-$B$-recurring infinitary vertex. By K\"{o}nig's lemma \cite{koening-theorie-der-endlichen-und-unendlichen-graphen-1936}, there is an infinite path starting in it. Assume that every $B$-tagged strict descendant of the vertex is finitary. Then, after removing the finitary vertices, each successor of the vertex is $B$-free, but the infinite path is still there and all of its vertices (except, maybe, the first one) are removed in the second step, decreasing the ultimate width by one. If there is a strict $B$-tagged infinitary descendant of the vertex, apply the same argument to it. This cannot go ad infinitum, because a path with an infinite number of $B$-tagged vertices would be constructed. 

This all implies:

\begin{lemma}[peeling \cite{kupferman-vardi-weak-alternating-automata-are-not-that-weak-2001}]
 For every B\"{u}chi automaton with $n$ states, peeling the run DAG of any $\omega$-word $n$ times yields the subgraph induced by the $B$-recurring vertices.\qed
\end{lemma}

\index{peeling!of a leveled DAG}

This can be used in various ways, in particular, it can be used for complementing B\"{u}chi automata, see Section~\ref{sec:complementation}, determinizing them backward, see Section~\ref{sec:backward-determinization}, and showing that alternating B\"{u}chi automata can easily be converted into weak alternating automata, see Section~\ref{sec:weak-alternation}.

To describe these applications, it is useful to have some notation and terminology at hand. By the above, each vertex $v$ in a $\{B\}$-tagged DAG of finite width can be assigned a value in $\omega \cup \{\infty\}$ according to when the vertex is removed by peeling the DAG successively. More precisely, when $i$ is a natural number and all vertices with value $< 2i$ are removed from the given DAG, the finitary vertices in the remaining DAG get assigned~$2i$; when all vertices with value $< 2i+1$ are removed, the $B$-free vertices in the remaining DAG get assigned~$2i+1$. The $B$-recurring vertices get assigned $\infty$. The number assigned to a vertex $v$ is called its canonical rank, it is denoted $c(v)$, and, according to the above, it is $\infty$ or $< 2n$, when $n$ is the width of the DAG.\index{canonical rank function}\index{rank function!canonical}

\begin{corollary} 
 \label{three}
 For a B\"{u}chi automaton with $n$ states, let $c$ be the canonical rank function of the run DAG of some $\omega$-word. 
 \begin{inparaenum}
 \item The word is accepted if, and only if, $c(v) = \infty$ for some $I$-tagged vertex $v$ on level~$0$. 
 \item Equivalently, the word is not accepted if, and only if, $c(v) < 2n$ for every $I$-tagged vertex $v$ on level~$0$.
 \end{inparaenum}
\end{corollary}

\subsection{Complementation via canonical ranks}
\label{sec:complementation}

The idea of using ranks or ``progress measures'' for complementing $\omega$-automata goes back to \cite{klarlund-progress-measures-complementation-omega-automata-applications-temporal-logic-1991} and has been improved and refined over the years, especially in \cite{kupferman-vardi-weak-alternating-automata-are-not-that-weak-2001}. The basic idea is to implement Corollary~\ref{three}(2). The starting point is a compilation of properties of the canonical rank function of a given $\{I, B\}$-tagged leveled DAG.

\begin{ennote}
 \label{one} 
 Let $v$ be any vertex. If $v$ does not have any successor, let $M = 0$, else let~$M$ be the maximum of all values $c(v')$ for successors~$v'$ of~$v$. If $v$ is not $B$-tagged or if $M$ is even, then $c(v) = M$; if $v$ is $B$-tagged and $M$ is odd, then $c(v) = M + 1$.
\end{ennote}

\begin{ennote}
 \label{two} 
 For any vertex with an even rank, the number of its descendants with the same rank is finite.
\end{ennote}

In general, a rank function of a leveled DAG of width~$n$ with vertex set~$V$ is a function $f \colon V \to [2n]$ satisfying Properties~\ref{one} and~\ref{two} with $f$ instead of $c$.\index{rank function!of a leveled DAG}\index{leveled DAG!rank function}

\begin{remark}
 Any rank function is pointwise greater or equal to the canonical rank function.
\end{remark}

A complementation construction for B\"{u}chi automata can now be based on Corollary~\ref{three}(2) and the following observations. First, there is a forward deterministic automaton with trivial recurrence condition that outputs the part of the $\{I, B\}$-tagged run DAG of a given word which is reachable from the $I$-tagged vertices on level~$0$. Second, there exists a nondeterministic B\"{u}chi automaton that produces for every $\{I, B\}$-tagged leveled graph of width at most~$n$ the same graph, but with any labeling with numbers from $[2n]$ such that Property~\ref{one} is satisfied. Third, using a variant of the breakpoint construction, see Theorem~\ref{thm:breakpoint}, a B\"{u}chi automaton can be constructed that checks Property~\ref{two} for a $[2n]$-labeled DAG. In other words, a suitable cascade yields a B\"{u}chi automaton for the complement of the language recognized by a given B\"{u}chi automaton.

\begin{theorem}[complementation via ranks
 \cite{kupferman-vardi-weak-alternating-automata-are-not-that-weak-2001}]
 \label{thm:kupferman-vardi}
 Complementation via canonical ranks yields, for every B\"{u}chi automaton with $n$ states, a B\"{u}chi automaton with at most $(6n)^n$ states.
\end{theorem}

In \cite{friedgut-kupferman-vardi-buechi-complementation-made-tighter-2006}, the above approach is improved, resulting in an asymptotic upper bound of $(0.96\, n)^n$ for the number of states, and in \cite{schewe-buechi-complementation-made-tight-2009} a further improvement leads to a construction which is optimal within a factor of $O(n^2)$ and has an asymptotic upper bound of $(0.76n)^n$.

\subsection{Backward determinization via canonical ranks}
\label{sec:backward-determinization}

A second application of canonical ranks is the conversion of a given nondeterministic B\"{u}chi automaton into an equivalent backward deterministic generalized transition-B\"{u}chi automaton. The idea, which is due to \cite{carton-michel-unambiguous-buechi-automata-2003}, is to use Corollary~\ref{three}(1) and to construct an automaton which labels the run DAG in a backward deterministic fashion with the values of the canonical rank function.

The key to designing such an automaton is the fact that the canonical rank function is the only function on an $\{I,B\}$-tagged DAG satisfying Property~\ref{one} and the following one, Property~\ref{twostar}.
\begin{ennote}
 \label{twostar}
 \begin{inparaenum}
 \item For every even number $i$, the set $c^{-1}(\{i\})$ is exactly the set of finitary vertices in the sub-DAG without the vertices in $c^{-1}([i])$.
 \item For every vertex $v \in V$, if $c(v) > 1$ and $c(v)$ is odd, then $v$ has a descendant $v'$ with $c(v') = c(v) - 1$.
 \item In the sub-DAG consisting of the vertices in $c^{-1}(\{\infty\})$, every vertex has a strict $B$-tagged descendant.
 \end{inparaenum}
\end{ennote}

This means a backward deterministic generalized transition-B\"{u}chi automaton computing the rank function for a run DAG can be constructed as a cascade of two automata: 
\begin{inparaenum}[(i)]
\item an automaton with a backward deterministic transition function and a trivial recurrence condition outputting the run DAG of a given word and an assignment to the vertices satisfying Property~\ref{one}; 
\item a backward deterministic automaton checking Property~\ref{twostar} using adaptations of the lift construction, see Theorem~\ref{thm:shift} and also the subsequent remark on weakly finitary vertices.
\end{inparaenum}
An automaton equivalent to the given B\"{u}chi automaton is obtained when the states which assign $\infty$ to an $I$-tagged vertex are chosen to be initial. 

\begin{theorem}[backward determinization via canonical ranks
 \cite{carton-michel-unambiguous-buechi-automata-2003}]
 Backward determinization via canonical ranks yields, for every B\"{u}chi automaton with $n$~states, a generalized transition-B\"{u}chi automaton with at most $(3n)^n$ states.
\end{theorem}

\section{Run trees of B\"{u}chi automata}
\label{run trees}

Run DAG's are one way to represent the set of all runs of a B\"{u}chi automaton on an $\omega$-word. A different approach, which can serve as a basis for complementation, disambiguation, and forward determinization, is to use compressed run trees.

In a first step towards the definition of the compressed run tree for a given word $u$ with respect to a given B\"{u}chi automaton, a labeled binary tree $t$ is defined, using a refined subset construction. Just as in the subset construction, all the states reachable from the initial states are tracked at the same time. The difference is that in each step the set of states reachable by reading the next letter is split into the B\"{u}chi states and the non-B\"{u}chi states: a binary tree emerges. To keep this tree compact, only one occurrence of each state---more precisely, its leftmost occurrence---is kept, that is, the tree is pruned in a straightforward fashion.

In the following, when a vertex $v$ is said to be to the left of another vertex $v'$, then this means that $v$ and $v'$ are on the same level, that is, $|v| = |v'|$, and there exists $i < |v|$ such that $v(j) = v'(j)$ for all $j<i$, $v(i) = 0$, and $v'(i) = 1$. The corresponding ordering is denoted by $\LeftLt$.

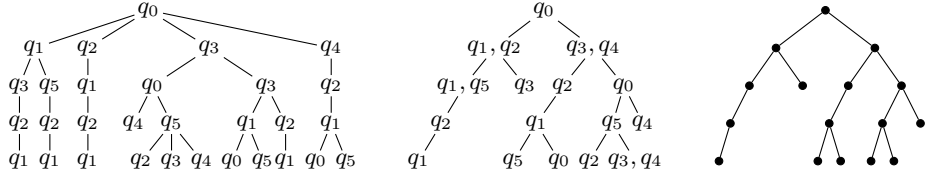
\begin{figure}
 \centering
 \small
 \begin{tikzpicture}[%
	 level distance=5mm,
	 level 1/.style={sibling distance=13mm},
	 level 2/.style={sibling distance=7mm},
	 level 3/.style={sibling distance=5mm},
	 level 4/.style={sibling distance=3mm}]

	 \begin{scope}[every node/.style={draw,circle,fill,inner sep=1pt}]
	 \node {}
		 child {node {} 
			 child {node {}
				 child {node {}
					 child {node {}}
					 child[missing]}
				 child[missing]}
			 child {node {}
			 }}
		 child {node {} 
			 child {node {}
				 child {node {}
					 child {node {}}
					 child {node {}}}
				 child[missing]
			 }
			 child {node {}
				 child {node {}
					 child {node {}}
					 child {node {}}} 
				 child {node {}}}};
	 \end{scope}

	 \begin{scope}[every node/.style={inner sep=1pt},
		 level distance=5mm,
		 level 1/.style={sibling distance=13mm},
		 level 2/.style={sibling distance=8mm},
		 level 3/.style={sibling distance=6mm},
		 level 4/.style={sibling distance=6mm}]
		 \node at (-3.7,0) {$q_0$}
		 child {node {$q_1, q_2$} 
			 child {node {$q_1, q_5$}
				 child {node {$q_2$}
					 child {node {$q_1$}}
					 child[missing]}
				 child[missing]}
			 child {node {$q_3$}
			 }}
		 child {node {$q_3, q_4$} 
			 child {node {$q_2$}
				 child[sibling distance=7mm] {node {$q_1$}
					 child {node {$q_5$}}
					 child {node {$q_0$}}}
				 child[missing]
			 }
			 child {node {$q_0$}
				 child[sibling distance=3mm] {node {$q_5$}
					 child {node {$q_2$}}
					 child {node {$q_3, q_4$}}} 
				 child[sibling distance=5mm] {node {$q_4$}}}};
	 \end{scope}

	 \begin{scope}[every node/.style={inner sep=1pt},
				 level 1/.style={sibling distance=16mm},
				 level 2/.style={sibling distance=15mm},
				 level 3/.style={sibling distance=5mm},
				 level 4/.style={sibling distance=4mm}]
	 \node at (-8.9,0) {$q_0$}
		 child[sibling distance = 10mm] {node {$q_1$} 
			 child[sibling distance = 4mm] {node {$q_3$}
				 child {node {$q_2$}
					 child {node {$q_1$}}}}
			 child[sibling distance = 4mm] {node {$q_5$} 
				 child {node {$q_2$}
					 child {node {$q_1$}}}}} 
		 child {node {$q_2$} 
			 child {node {$q_1$}
				 child {node {$q_2$}
					 child {node {$q_1$}}}}} 
		 child {node {$q_3$} 
			 child {node {$q_0$}
				 child {node {$q_4$}}
				 child {node {$q_5$}
					 child {node {$q_2$}}
					 child {node {$q_3$}}
					 child {node {$q_4$}}}} 
			 child {node {$q_3$}
				 child {node {$q_1$}
					 child {node {$q_0$}}
					 child {node {$q_5$}}}
				 child {node {$q_2$}
					 child {node {$q_1$}}}}} 
		 child {node {$q_4$} 
			 child {node {$q_2$}
				 child {node {$q_1$}
					 child {node {$q_0$}}
					 child {node {$q_5$}}}}} ;
	 \end{scope}
 \end{tikzpicture}
 \caption{The first five levels of the ordinary run tree of a word; the corresponding part of the labeled compressed run tree; the corresponding part of the compressed run tree. For ease in notation, curly brackets to denote sets are omitted.}
 \label{fig:run-tree}
\end{figure}

The definition of $t$ is by induction on the levels. The root (level~$0$) of $t$ is labeled with~$Q_I$. Assume all vertices on level $l$ have already been constructed and assigned labels, say these vertices form the set $W$, and let $a$ stand for the next letter, $u(l)$. For each $v \in W$, let $R_v = \bigcup \{\Delta(q,a) \mid q \in t(v)\}$, which means $R_v$ is the set of states reached from any state in the label of $v$ by reading~$a$. Set
\begin{align}
 Q_v^0 & = (R_v \cap B) \setminus  \bigcup_{w \LeftLt v} R_w\enspace,&
 Q_v^1 & = R_v  \setminus (B \cup \bigcup_{w \LeftLt v} R_w) \enspace.
\end{align} 
In other words, $Q_v^0$ is the set of B\"{u}chi states reached from $t(v)$ by reading letter~$a$, but not including the states that are reached from any state in a label of a vertex to the left of~$v$. Similarly, $Q_v^1$ is the set of non-B\"{u}chi states reached from $t(v)$ by reading letter~$a$, but not including the states that are reached from any state in a label of a vertex to the left of~$v$. The definition of the tree $t$ now says that, for $v \in W$ and $i<2$, if $Q_v^i \neq \emptyset$, then $vi \in t$ and $t(vi) = Q_v^i$. For an illustration, see Figure~\ref{fig:run-tree}. 

The tree $t$ is called the labeled compressed run tree of~$u$, while the unlabeled tree, that is, the domain of $t$, is called the compressed run tree of~$u$ and denoted~$T$.\index{labeled compressed run tree}\index{run tree!labeled compressed}

\begin{lemma} 
 \label{thm:run-trees}
 A B\"{u}chi automaton accepts an $\omega$-word if, and only if, its compressed run tree has a path with infinitely many left successors, also called left-recurring path.\index{left-recurring!path}\index{path!left-recurring}
\end{lemma}

For the proof, assume $u$ is a word which is accepted by a given B\"{u}chi automaton. We construct an infinite path $\langle v_0, v_1, \dots\rangle$ in $t$ in such a way that for every $l$ the following conditions hold.
\begin{inparaenum}[(i)]
\item There exists a recurring run on $u[l,\infty)$ starting in some $q \in t(v_l)$.
\item There is no vertex $v \LeftLt v_l$ satisfying the same condition.
\end{inparaenum}
For the induction base, we choose $v_0 = \epsilon$, which obviously works. Assume $v_l$ has already been defined. By~(i), there is a recurring run on $u[l,\infty)$ starting in some state of $t(v_l)$. If there is such a run~$r$ with $r(1) \in B$, we set $v_{l+1} = v_l 0$ or else $v_{l+1} = v_l 1$. To show $v_{l+1} \in T$, we fix a state $q \in t(v_l)$ and a recurring run $r$ on $u[l+1,\infty)$ such that $qr$ is a recurring run on $u[l,\infty)$. By way of contradiction, assume $v_{l+1} \notin T$. Then there are a vertex $v \LeftLt v_l$ and a state $q' \in t(v)$ such that $r(0) \in \Delta(q', u(l))$. This means $q'r$ is a recurring run on $u[l,\infty)$---a~contradiction to~(ii). 

Assume the constructed path is not left-recurring. Then there exists $i$ such that $v_j \in (0+1)^*1$ for all $j \geq i$. Let $r$ be a recurring run on $u[i,\infty)$ starting with a state in $t(v_i)$. Then there is some $k>0$ such that $r(k)$ is a B\"{u}chi state. By adjusting $i$, we can assume $k=1$. If, on one hand, $r(1) \notin t(v_{i+1})$, we can obtain a contradiction similar to above. If, on the other hand, $r(1) \in v_{i+1}$, then, by definition, $v_{i+1} = v_{i+1-1}0$---a contradiction.

For the converse, assume $\langle v_0, v_1, \dots\rangle$ is a left-recurring path in $T$. For every $i < \omega$ and every $q \in t(v_i)$, there is an initial run of the automaton on $u[0,i)$ such that $r(i) = q$ and $r(j) \in t(v_j)$ for every $0 < j < i$. All these runs can be organized in a straightforward fashion in an infinite tree with branching degree at most the number of states of the given automaton. By K\"{o}nig's lemma, this tree has an infinite rooted path, and, by construction, the labeling of this path is an initial run of the automaton on~$u$. Further, for every $i>0$, the state in position~$i$ of this run belongs to~$B$ if, and only if, $v_i$ is a left successor.\qed

From an automata-theoretic point of view the important observation is that compressed run trees of a given B\"{u}chi automaton have width at most $n$ and can be constructed in a forward deterministic fashion by an $\omega$-automaton with output and trivial recurrence condition. One way to realize such an automaton is to use states of the form $\langle Q_0, \dots, Q_{m-1}\rangle$ with the $Q_i$'s being pairwise disjoint, nonempty subsets of $Q$, representing the labeling of the current level of the labeled compressed run tree. An upper bound on the number of such states can be obtained using ordered Bell numbers.

\begin{remark}
 \label{thm:run-trees-deterministic}
 For every B\"{u}chi automaton with $n$ states, there is a forward deterministic automaton that outputs, for every $\omega$-word, a representation of its compressed run tree, and has a number of states which is asymptotically bounded  from above by $(0.54 \, n)^n$.
\end{remark}

\subsection{Complementation and disambiguation via compressed run trees}
\label{sec:disambiguation}

Compressed run trees can be used for complementation. To see this, consider the subtree of a compressed run tree which contains only the infinitary vertices and call it the core of the run tree.\index{core!of a run tree}\index{run tree!core of} From Lemma~\ref{thm:run-trees} and the fact that the run tree has finite width it follows that an $\omega$-word is not accepted if, and only if, in the core of its run tree there are only finitely many slices with a left successor. In other words, a B\"{u}chi automaton for the complement is obtained as a cascade of the following automata: 
\begin{inparaenum}[(i)]
\item the automaton from Remark~\ref{thm:run-trees-deterministic}; 
\item an automaton based on the breakpoint construction removing the finitary vertices; 
\item a two-state automaton checking that from some level onward, no slice with a left successor occurs anymore.
\end{inparaenum}
A careful implementation leads to state spaces similar in size to those described in~\cite{schewe-buechi-complementation-made-tight-2009}.

An interesting observation is that the cascade of the first and the second automaton from above yields a nondeterministic B\"{u}chi automaton which, for every $\omega$-word over the given alphabet, outputs the core of its compressed run tree and has exactly one accepting run. In general, an automaton which has at most one accepting run for each word is called an unambiguous automaton.\index{unambiguous!$\omega$-automaton}\index{$\omega$-automaton!unambiguous} In other words, the above automaton is an unambiguous automaton for the set of all $\omega$-words over the given alphabet. Moreover, it can be modified in two ways.
\begin{inparaenum}[(i)]
\item By cascading it with a two-state deterministic automaton checking
 that there are infinitely many slices with left successors, one
 obtains an unambiguous automaton for the language recognized by the given automaton.
\item By cascading it with a two-state unambiguous automaton checking that there are only finitely many slices with left successors, one obtains an unambiguous automaton for the complement.
\end{inparaenum}
In the terminology of~\cite{colcombet-forms-determinism-automata-2012}, such an automaton could be called strongly unambiguous.\index{strongly unambiguous!$\omega$-automaton}\index{$\omega$-automaton!strongly unambiguous}






\subsection{Forward determinization via compressed run and history trees}

Theorem~\ref{thm:fundamental} states in particular that every nondeterministic B\"{u}chi automaton is equivalent to a forward deterministic parity, Rabin, Streett, or Muller automaton. The quest for good constructions establishing this---determinization constructions---has resulted in different approaches. The approach followed in this section is motivated by \cite{muller-schupp-alternating-automata-infinite-trees-1987}, but it is also closely related to Safra-like constructions, as explained towards the end.

In view of Lemma~\ref{thm:run-trees} and Remark~\ref{thm:run-trees-deterministic}, a determinization construction has been established once it has been shown that the set of all binary trees of width at most $k$ which have a left-recurring path is recognized by a forward deterministic automaton. Therefore, the objective in what follows is exactly to describe such an automaton.

\begin{figure}[t]
 \centering
 \small
 \begin{tikzpicture}[%
	 level distance=5mm,
	 level 1/.style={sibling distance=13mm},
	 level 2/.style={sibling distance=7mm},
	 level 3/.style={sibling distance=5mm},
	 level 4/.style={sibling distance=3mm}]

	 \begin{scope}[every node/.style={draw,circle,fill,inner sep=1pt}]
	 \node {}
		 child {node {} 
			 child {node {}
				 child {node {}
					 child {node {}}
					 child[missing]}
				 child[missing]}
			 child {node {}
			 }}
		 child {node {} 
			 child {node {}
				 child {node {}
					 child {node {}}
					 child {node {}}}
				 child[missing]
			 }
			 child {node {}
				 child {node {}
					 child {node {}}
					 child {node {}}} 
				 child {node {}}}};
	 \end{scope}

	 \begin{scope}[every node/.style={draw=none,fill=white,inner sep=1pt}]
	 \node (o3) at (4,0) {0}
		 child {node (o2) {1} 
			 child {node {2}
				 child {node {1}
					 child {node (l0) {1}}
					 child[missing]}
				 child[missing]}
			 child {node {1}
			 }}
		 child {node {0} 
			 child {node (o4) {3}
				 child {node {2}
					 child {node (l1) {3}}
					 child {node (l2) {2}}}
				 child[missing]
			 }
			 child {node {0}
				 child {node {3}
					 child {node (l3) {4}} 
					 child {node (l4) {0}}} 
				 child {node {0}}}};
		 \draw[densely dashed,->] (l2) .. controls +(45:0.5) and +(300:0.5) .. (o4);
		 \draw[densely dashed,->] (l4) .. controls +(0:1.5) and +(0:1.5) .. (o3);
		 \draw[densely dashed,->] (l1) .. controls +(240:0.7) and +(300:0.7) .. (l1);
		 \draw[densely dashed,->] (l3) .. controls +(240:0.7) and +(300:0.7) .. (l3);
		 \draw[densely dashed,->] (l0) .. controls +(170:0.9) and +(190:0.9) .. (o2);
	 \end{scope}

	 \begin{scope}[every node/.style={draw,circle,fill,inner sep=1pt}]
	 \node at (8,0) {}
			child {node {}}
			child {node {}
				child {node {}}}
			child {node {}};
		\end{scope}
 \end{tikzpicture}
 \caption{The first five levels of a compressed run tree; the same tree with the origin numbers of each vertex and the origins of the vertices on level $4$; the tree induced by the origins of the vertices on level~$4$. There are two origins that move left in the last slice depicted.}
 \label{fig:origin}
\end{figure}
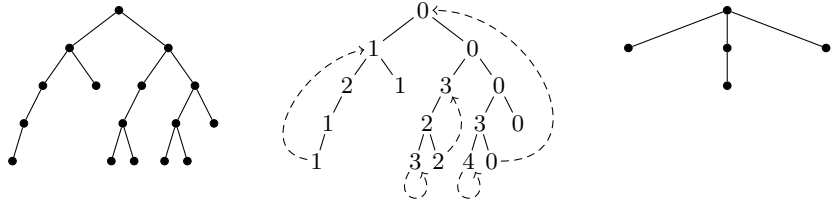

To understand the mechanics of infinite trees of finite width better, we associate with every vertex $v$ in a binary tree its origin.\index{origin!of a vertex in a binary tree} This is the earliest  ancestor of~$v$ (shortest prefix of $v$) with the property that no vertex to the right of $v$ has the same ancestor; it is denoted $\text{or}(v)$. For an illustration, see Figure~\ref{fig:origin}. 

Observe, for instance, that
\begin{inparaenum}[(i)]
\item the root is the origin of the rightmost vertex on each level; 
\item vertices on the same level have distinct origins; 
\item if a vertex has a left and a right successor, then the left successor is its own origin.
\end{inparaenum}

The important definition specifies that an origin moves left in one slice if it is the origin of a vertex $v$ on the upper level of the slice and of a vertex $v'$ on the lower level of the slice and $v'$ is not the right successor of $v$, that is, $v'$ is to the left of $v1$. (Note that, by definition, $v'$ cannot be to the right of $v1$.) For an illustration, see Figure~\ref{fig:origin}.

The key for the construction to be presented is:

\begin{lemma}
 A binary tree of finite width contains a left-recurring path if, and only if, there is some origin which moves left in infinitely many slices.
\end{lemma}

To prove the lemma, assume $\langle v_0, v_1, \dots\rangle$ is a left-recurring rooted path. Let $i$ be minimal such that there is no other infinite path $\langle v_0, \dots, v_j, v_j1, \dots\rangle$ for any $j \geq i$. (This number $i$ exists because there are at most $k$ rooted infinite paths in any tree of width $k$.) For every $j \geq i$, consider the rightmost vertex $w_j$ on level $j$ such that $v_j \LeftLeq w_j$ and there is no infinitary vertex $v'$ with $v_j \LeftLt v' \LeftLeq w_j$. Then $v_i$ is the origin of every $w_j$ and moves left in infinitely many slices. 

For the converse, let $v$ be an origin which moves left in infinitely many slices. Consider, for every level $i \geq |v|$, the vertex $w_i$ with origin $v$. By K\"{o}nig's lemma, there is a rooted path $\langle v_0, v_1, \dots\rangle$ in the tree which consists of all vertices $w_i$ and their ancestors. This path is left-recurring, because otherwise there would be some $j$ with $v_k \in v_j1^*$ for all $k \geq j$, which is a contradiction to $v$ moving left in infinitely many slices.\qed 

There are several ways for an automaton to check whether there is an origin which moves to the left in infinitely many slices.  One is explained in what follows, another one is sketched later.

We use the notion of military ordering, denoted $<_\text{mil}$, and defined by $v <_\text{mil} v'$ if, and only if, either $|v| < |v'|$ or $|v| = |v'|$ and $v \LeftLt v'$.\index{military ordering}\index{ordering!military}

From level to level, the automaton determines, for each vertex, its origin number,\index{origin number!of a vertex in a binary tree} which is defined as follows. For a given level $l$, let $v_0, \dots, v_{r-1}$ be an enumeration of the vertices on level~$l$, ordered according to the military order of their origins, that is,  $\text{or}(v_0) <_\text{mil} \dots <_\text{mil} \text{or}(v_{r-1})$. The index $i$ is the origin number of $v_i$ and denoted $\text{on}(v_i)$. The index $i$ is said to refer to the origin $\text{or}(v_i)$ on level $l$. For an illustration, see Figure~\ref{fig:origin}. 

When a forward deterministic automaton computes the origin numbers, it can be augmented by a transition-Rabin condition to check for a left-recurring path. To see this, assume $v$ is an origin on some level $l$ that moves infinitely often to the left and let $v_0, v_1, \dots$ be the list of vertices such that $v_i$ is on level $l+i$ and $\text{or}(v_i) = v$, in particular, $v_0 = v$. Then there is some $i$ such that $\text{on}(v_i) = \text{on}(v_{i+1}) = \dots$. So an appropriate transition-Rabin recurrence condition can be chosen to have, for each $i < k$ (maximum width of the trees considered), a Rabin pair $\langle L_i, U_i\rangle$ as follows. The set $L_i$ contains all transitions where for every $j \leq i$ the index $j$ does not refer to the same origin in the upper and the lower level of the current slice. The set $U_i$ contains all transitions where the origin which $i$ refers to on the upper level moves left in the slice.

As states of a forward deterministic automaton computing the origin numbers one can choose bijections $g \colon [m] \to [m]$ where $m \leq k$. The meaning of such a state $g$ would be that if $v$ is vertex $i$ on the current level in the order from left to right, then $\text{on}(v) = g(i)$. The actual definition of the transition function is somewhat technical and omitted. 

In the determinization constructions presented in \cite{safra-complexity-omega-automata-1988, Piterm2007a, schewe-tighter-bounds-for-determinization-buechi-automata-2009}, states are trees. For instance, in \cite{schewe-tighter-bounds-for-determinization-buechi-automata-2009}, so-called history trees\index{history tree} are used and defined as trees where
\begin{inparaenum}[(i)]
\item each node is labeled by a nonempty set of states, 
\item the label of every node is a strict superset of the union of the labels of its children, and 
\item the labels of siblings are disjoint.
\end{inparaenum}
Observe that, alternatively, one could require that the vertices of a history tree are labeled with pairwise disjoint, nonempty sets of states. Such a tree is obtained from the labeled compressed run tree of a given word for each level in a straightforward fashion: move every label of a vertex on the respective level to its origin and then remove all vertices except for these origins and contract edges accordingly. For an illustration, see Figure~\ref{fig:origin}.

If one constructs a deterministic automaton which only keeps track of the history trees just described, one arrives at a fundamental construction, which is also known to be optimal in a certain sense \cite{colcombet-zdanowski-tight-lower-bound-for-determinization-transition-labeled-buechi-automata-2009}:

\begin{theorem}[determinization via history trees
 \cite{schewe-tighter-bounds-for-determinization-buechi-automata-2009}]
 \label{thm:schewe}
 Determinization via history trees yields, for every B\"{u}chi automaton with $n$ states, an equivalent deterministic transition-Rabin automaton with an asymptotic upper bound of $(1.65\,n)^n$ for the number of states and $2^n-1$ Rabin pairs.
\end{theorem}

\enlargethispage{0\baselineskip}

The transformations on different types of $\omega$-automata discussed in this section and the previous one are fundamental transformations, but not all one can consider. Optimal solutions for most of the basic transformation tasks can be found in \cite{safra-complexity-omega-automata-1988} and a later paper by the same author \cite{safra-exponential-determinization-for-omega-automata-with-a-strong-fairness-acceptance-condition-2006}. Here, ``optimal'' means with respect to a rough measure of complexity: polynomial, exponential, doubly exponential. The ``optimal'' results stated subsequent to Theorem~\ref{thm:kupferman-vardi} and before Theorem~\ref{thm:schewe} are with respect to much finer measures and based on very good lower bounds. A breakthrough with regard to lower bounds on $\omega$-automata is \cite{yan-lower-bounds-complementation-omega-automata-full-automata-technique-2006}.

\section{Congruence relations}

Congruence relations are a useful tool for working with finite-state automata on finite words. For instance, the minimum-state deterministic finite-state automaton for a given regular language of finite words can be derived from the Myhill-Nerode congruence relation for the language---it merely is this congruence relation.

For $\omega$-automata, congruences are also useful, but the situation is more complex. 

\subsection{Right (and left) congruence relations}
\label{sec:finite-state}

The straightforward adaptation of the Myhill-Nerode congruence relation (see, for instance, \cite{MN}) to $\omega$-lan\-guages is the initial syntactic congruence relation.\index{initial syntactic congruence relation!of an $\omega$-language}\index{congruence relation!initial syntactic! of an $\omega$-language} For a given $\omega$-lan\-guage~$L$, it considers finite words $u$ and $v$ equivalent if, and only if, for every  $\omega$-word $w$, either $\{uw, vw\} \subseteq L$ or $\{uw,vw\} \cap L = \emptyset$. 

On the automata-theoretic side, there is a corresponding notion. Given an $\omega$-au\-tom\-a\-ton of any type, its initial congruence relation\index{initial congruence relation!of an $\omega$-automaton}\index{congruence relation!initial! of an $\omega$-automaton} considers finite words $u$ and $v$ equivalent if, and only if, for every initial run of the automaton on~$u$ ending in some state~$q$ there is such a run on~$v$, and vice versa.

The analogy to the Myhill-Nerode congruence relation is as follows.

\begin{remark}
 \begin{inparaenum}
 \item The initial syntactic congruence relation for an $\omega$-language is a right congruence relation. 
 \item For an $\omega$-automaton of any type, the initial congruence relation is a right congruence relation and equal to or finer than the initial syntactic congruence relation for the language recognized and has a finite number of congruence classes.
 \end{inparaenum}
\end{remark}

(Here, as usual, an equivalence relation on words over a given alphabet is a right congruence relation if $uw$ and $vw$ are equivalent whenever $u$ and $v$ are and $w$ is a finite word over the alphabet. A relation is finer than another one if it is a subset of it.)

For a simple language such as the set of all ultimately periodic words over a given alphabet, which is not regular, the initial syntactic congruence relation has only one equivalence class. Hence, it cannot serve as a vehicle to define regularity. It can neither be used for classifying regular $\omega$-languages: the languages denoted by $(0+1)^\omega$ and by $(0^*1)^\omega$ have the same initial syntactic congruence relation, but they are completely different in nature.

The initial syntactic congruence relation can, however, take over the role of the Myhill-Nerode congruence relation for the small class of $\omega$-languages which are recognized by forward deterministic weak automata (see also Section~\ref{thm:weak-characterization}):

\begin{theorem}[minimization of forward deterministic weak automata
 \cite{staiger-finite-state-omega-languages-1983,loeding-efficient-minimization-deterministic-weak-omega-automata-2001}]
 \label{thm:initial-congruence} 
 Let $L$ be an $\omega$-language recognized by a forward deterministic weak automaton $\Aut A$ and let $\Aut D$ be the DFA (without final state set) corresponding to the initial syntactic congruence relation for~$L$.
 \begin{enumerate}
 \item The automaton $\Aut D$ can be augmented by a weak acceptance condition in such a way that the resulting automaton recognizes $L$.

 \item The automaton from (1) is, up to isomorphism, the smallest forward
 deterministic automaton recognizing~$L$ and can be computed from $\Aut A$ by
 DFA minimization, see, for instance, \cite{HMU}.
 \end{enumerate}
\end{theorem}

The role that the initial syntactic congruence relation of a language~$L$ plays in the context of forward deterministic automata is taken over by the final syntactic congruence relation in the context of backward deterministic automata. This relation, which is a left congruence relation, considers $\omega$-words $v$ and $w$ congruent if, and only if, for every  finite word $u$, either $\{uv, uw\} \subseteq L$ or $\{uv,uw\} \cap L = \emptyset$.

\subsection{Two-sided congruence relations}
\label{sec:saturation}

There are essentially two straightforward adaptations of the two-sided syntactic
congruence relation for languages of finite words (see, for instance, \cite{SC}) to $\omega$-languages.

Let $L$ be an $\omega$-language over some alphabet $A$. In the first adaptation, finite words $u$ and $v$ are congruent if, and only if, for all $x \in A^*$ and $y \in A^\omega$, either $\{xuy, xvy\} \subseteq L$ or $\{xuy, xvy\} \cap L = \emptyset$. In the second adaptation, nonempty finite words $u$ and $v$ are congruent if, and only if, whenever $u_0, u_1, \dots$ and $v_0, v_1, \dots$ are sequences of nonempty finite words over the given alphabet such that $u_i = v_i$ or $\{u_i, v_i\} \subseteq \{u,v\}$, then either $\{u_0 u_1 \dots, v_0 v_1 \dots\} \subseteq L$ or $\{u_0 u_1 \dots, v_0 v_1 \dots\} \cap L = \emptyset$.

The two adaptations try to capture what it means for two finite words to behave equally in the same context, and they both yield two-sided congruence relations. The first one is finer or equal to the initial syntactic congruence relation; the second one is finer or equal to the first one and called the syntactic congruence relation of~$L$.\index{syntactic congruence relation!of an $\omega$-language}\index{congruence relation!syntactic! of an $\omega$-language}

In the following, the syntactic congruence relation is further discussed, because it provides a finer means of characterization.

Corresponding to the syntactic congruence relation one can define, for every $\omega$-au\-tom\-a\-ton of any type, a suitable two-sided congruence relation. For a B\"{u}chi automaton, this relation considers nonempty finite words $u$ and $v$ congruent if the following two conditions hold for all states $q, q' \in Q$.
\begin{inparaenum}[(i)]
\item There is a run of the automaton on $u$ starting with $q$ and ending in $q'$ if, and only if, this is true for $v$.
\item There is a run $r$ of the automaton on $u$ starting with $q$, ending in $q'$, and passing through an element of $B$, that is, $r(i) \in B$ for some $i < |r|$, if, and only if, this is true for~$v$.
\end{inparaenum}

Just as before, the syntactic congruence relation of an $\omega$-language does not characterize regularity. Consider%
\footnote{Slides of a presentation given by Miko\l aj Boja\'nczyk.} the language $L_\text{ub0}$ of all $\omega$-words of the form $0^{i_0}10^{i_1}\dots$ where $\limsup_{j \to \infty} i_j = \infty$. The language $L_\text{ub0}$ and the one denoted by $(0^*1)^\omega$ have the same syntactic congruence relation, and this has just two equivalence classes.

Still, the syntactic congruence relation has interesting properties. One is phrased in terms of saturation, where an equivalence relation on finite words is said to saturate \index{saturation!of an $\omega$-language} an $\omega$-language $L$ if, for all sequences $\langle V_0, V_1 \dots\rangle$ of equivalence classes, either $V_0V_1 \dots \subseteq L$ or $V_0 V_1 \dots \cap L = \emptyset$ holds.

\begin{theorem}[saturation \cite{buechi-decision-method-restricted-second-order-arithmetic-1962,arnold-syntactic-congruence-omega-languages-1983}]%
 \label{thm:syntactic-congruence}
 \begin{enumerate}
	 \item The two-sided au\-tom\-a\-ton congruence relation of a B\"{u}chi automaton with $n$ states has at most $2^{2n^2}$ congruence classes and saturates the language recognized by the automaton.
 \item The syntactic congruence relation of a given regular $\omega$-language is the coarsest congruence relation saturating the language.
 \item An $\omega$-language is regular if, and only if, there exists a congruence relation saturating it and having a finite number of congruence classes.
 \end{enumerate}
\end{theorem}

The proofs of (1) and (2) are straightforward; one direction of~(3) follows from~(1). The other direction of~(3) can be proved on the basis of Ramsey's Theorem~A\index{Ramsey's Theorem A} \cite{ramsey-problem-formal-logic-1929}, which says that, for a given equivalence relation on finite words with a finite number of equivalence classes and an infinite sequence of nonempty finite words $u_0,u_1, \dots$, there is a strictly monotone infinite sequence $\langle i_0, i_1, \dots\rangle$ of natural numbers such that all finite words of the form $u_{i_j} u_{i_j+1} \dots u_{i_k-1}$ with $j < k$ are equivalent. In the context of $\omega$-languages, this means the following.

\begin{remark}
 \label{thm:ramsey}
\cite{buechi-decision-method-restricted-second-order-arithmetic-1962}
 Given an alphabet $A$, a congruence relation on $A^+$ having a finite number of congruence classes, and $u \in A^\omega$, there are congruence classes $U$ and $V$ satisfying $UV \subseteq U$, $V^2 \subseteq V$, and $u \in UV^\omega$.
\end{remark}

To prove the other direction of Theorem~\ref{thm:syntactic-congruence}(3), observe that from the previous remark it follows that if a congruence relation with a finite number of equivalence classes saturates a given $\omega$-language~$L$, then $ L = \bigcup UV^\omega$ with $U$ and $V$ ranging over equivalence classes with \mbox{$UV^\omega \subseteq L$}. Based on this, a construction such as the one described in the proof of Theorem~\ref{thm:regexps} can be used to arrive at a B\"{u}chi automaton recognizing~$L$.\qed

The procedure just described is far from being as natural as the procedure that turns the Myhill-Nerode congruence relation for a given regular language of finite words into the minimum-state DFA for the language. In fact, nothing which would come close to this is known for regular $\omega$-languages. Still, two-sided congruence relations for $\omega$-languages are useful in several contexts, for instance, when it comes to classifying regular $\omega$-languages, see~\cite{perrin-pin-infinite-words-2004}. Another application, described in what follows, is complementation.

\subsection{Complementation via two-sided congruence relations}

When a two-sided congruence relation saturates a given language, then, by definition, it also saturates the complement of the language, which establishes once again that the class of $\omega$-languages recognized by B\"{u}chi automata is closed under complementation. In fact, the first proof of this fact was along these lines \cite{buechi-decision-method-restricted-second-order-arithmetic-1962}.

In view of the bound stated in Theorem~\ref{thm:syntactic-congruence}(1), using congruences for complementation leads to much larger automata than the ones described in Sections~\ref{sec:complementation} and~\ref{sec:disambiguation}. To obtain smaller automata with this approach, the following modification suggests itself. 

Write the complement of the language recognized by a given B\"{u}chi automaton again in the form $\bigcup_{i < k} U_i V_i^\omega$, but choose the $U_i$'s and $V_i$'s to be unions of congruence classes in a way such that the resulting B\"{u}chi automaton is small. Observe that if $m$ is the number of states of an automaton recognizing all $U_i$'s (with a different set of final states for each~$i$) and $m'$ is an upper bound on the number of states needed in automata recognizing the $V_i$'s, then $m + k (m' +1)$ is an upper bound for the number of states in the resulting B\"{u}chi automaton.

To see how the indicated approach works, assume a B\"{u}chi automaton is given as usual, with $n$ states and recognizing a language $L$. For every set $P \subseteq Q$, let $U_P$ be set of words $u$ such that every run of the automaton on $u$ starting in some initial state ends in some state of~$P$. Then each set $U_P$ is a union of equivalence classes of the automaton congruence relation and can be recognized by a deterministic automaton with $2^n$ states.

For every nonempty sequence $\sigma = \langle P_0, \dots, P_{k-1}\rangle$ of nonempty, pairwise disjoint sets of states, let the set $P(\sigma)$ be defined by $P(\sigma) = P_0 \cup \dots \cup P_{k-1}$, and let $V_\sigma$ be the set of all finite words $u$ satisfying the following two conditions for every run $r$ of the automaton on~$u$.
\begin{inparaenum}[(i)]
\item If $r(0) \in P_i$ for some $i$, then $r(|u|) \in P_j$ for some $j$ with $i \leq j < k$.
\item If $r(0) \in P_i$ and $r$ contains a B\"{u}chi state, then $r(|u|) \in P_j$ for some $j$ with $i < j < k$. 
\end{inparaenum}

\begin{theorem}[complementation by saturation \cite{breuers-loeding-olschewski-improved-ramsey-based-complementation-2012}]
 For every B\"{u}chi automaton with $n$ states, the language $\bigcup_\sigma U_{P(\sigma)} V_\sigma^\omega $
 is the complement of the language recognized by the B\"{u}chi automaton, and a conversion of this expression into a B\"{u}chi automaton yields an automaton with $2^{\theta(n \log n)}$ states.
\end{theorem}

For the proof, first observe that each set $V_\sigma$ is a union of equivalence classes of the two-sided automaton congruence relation: compare (i) and (ii) above with (i) and (ii) in the definition of the two-sided automaton congruence relation. 

Next, let $\sigma$ and $P(\sigma)$ be as above. To see that $U_{P(\sigma)} V_\sigma^\omega \cap L = \emptyset$ holds, let $r$ be an initial run on any word $u \in U_{P(\sigma)} V_\sigma^\omega$ and $i_0 < i_1 < \dots$ be such that $u[0,i_0) \in U_{P(\sigma)}$ and $u[i_j, i_{j+1}) \in V_\sigma$ for every $j$. From the definition of $V_\sigma$ we can conclude that $r(i_0) \in P(\sigma)$ holds and that there are $i$ and $k$ such that $r(i_j) \in P_i$ holds for every $j \geq k$. This implies that, for every $j \geq k$, there is no B\"{u}chi state in $r[i_j, i_{j+1})$, which means $u \notin L$.

Conversely, if an $\omega$-word $u$ is not accepted by the given B\"{u}chi automaton, then, by Remark~\ref{thm:ramsey}, there are classes $U$ and $V$ of the automaton congruence relation such that $u \in UV^\omega$, $UV \subseteq U$, and $V^2 \subseteq V$. Let $P$ be the set of states which can be reached by reading some word from $U$ from some initial state. Consider the graph with vertex set~$P$ and an edge between $q$ and $q'$ if, and only if, there is a run of the automaton on some word $v \in V$ starting in $q$ and ending in $q'$. Let $\sigma = \langle P_0, \dots, P_{k-1} \rangle$ be a list of the SCC's of this graph in topological order. Then $V \subseteq V_\sigma$, which means $u \in U_{P(\sigma)} V_\sigma^\omega$. 

To prove the claim about the size of the resulting $\omega$-automaton, we describe how to construct a deterministic automaton of size $(k+1)^n$ for a language $V_\sigma$ as above. The states are functions $Q \to \{-\infty\} \cup [k]$. The transition function is defined in a way such that if by reading a finite word $v$ the automaton reaches state $f$, then the following holds for every $q \in Q$. If in the B\"{u}chi automaton there is no run on $v$ starting in $P(\sigma)$ and ending in $q$, then $f(q) = - \infty$; if there are such runs, then $f(q)$ is the greatest index $i$ such that a run on $v$ starting in some state from $P_i$ ends in $q$.\qed

The construction described above can be generalized so as to improve the construction of a B\"{u}chi automaton from a saturating congruence relation.

\section{Loop structure}

As the set of states visited infinitely often in a run of an $\omega$-automaton determines whether the run is recurring, it is only natural to investigate the structure of the strongly connected subsets in a given $\omega$-automaton.

A loop\index{loop!in an $\omega$-automaton} at some state $q$ is a word $q_0 a_0 q_1 a_1 \dots a_n q_{n+1}$ where $\langle q_i, a_i, q_{i+1}\rangle \in \Delta$ for every $i \in [n]$ and $q_0 = q_{n+1} = q$. The word $a_0 \dots a_n$ is the label of the loop, the set $\{q_0, \dots, q_n\}$ is the loop set. The loop is positive if it satisfies the recurrence condition of the given automaton (for a B\"{u}chi condition, this means $\{q_0, \dots, q_n\} \cap B \neq \emptyset$), it is negative if it does not---we speak of the sign of the loop. In a deterministic automaton (forward or backward), $q$ and the label determine the loop.

In forward deterministic automata, the nesting depth of positive and negative loops sets is an interesting measure for the complexity of the language recognized, explained in  Section~\ref{sec:wagner}, whereas in backward deterministic automata, the distribution of the labels of positive loops is interesting, as explained in Section~\ref{sec:backward-loops}. 

In the following, we assume, without loss of generality, that in forward deterministic automata every state is reachable from the initial state.

\subsection{Alternating loops in forward deterministic automata}
\label{sec:wagner}

A tower\index{tower!in an $\omega$-automaton} is a nonempty sequence $\langle C_0, \dots, C_{m-1}\rangle$ of loop sets such that $C_0 \supseteq \dots \supseteq C_{m-1}$ and the signs alternate; the sign of the last loop is the sign of the tower, the number~$m$ is the height of the tower. A maximal tower is one of maximal height. A wall\index{wall!in an $\omega$-automaton} is a sequence of maximal towers where each one is reachable from the previous one and the signs alternate; the sign of the wall is the sign of the first tower, the number of towers in the sequence is the length of the wall.

The types of towers and walls in a given forward deterministic $\omega$-automaton are invariants of the language recognized:

\begin{theorem}[towers and walls \cite{wagner-1977}]
 \label{thm:wagner}
 All forward deterministic $\omega$-automata recognizing the same language have the same types of towers and walls in the sense that if one of them has a tower of a certain height and sign or a wall of a certain length and sign, then the other has so, too. 
\end{theorem}

To illustrate this theorem we prove the claim for towers and start with a useful remark.

\begin{remark}
 \label{thm:loops}
 Consider a forward deterministic automaton with~$n$ states over an alphabet~$A$. For $u \in A^*$, $v \in A^+$ and $k \geq n$, some power of $v^k$ is the label of a loop at $\delta^*(q_I, uv^k)$, and this loop is positive if, and only if, $uv^\omega$ is accepted. (As usual, $\delta^*$ stands for the extended transition function, defined by $\delta^*(q, \epsilon) = q$ and $\delta^*(q, ua) = \delta(\delta^*(q,u),a)$ for all $q \in Q$, $u \in A^*$, and $a \in A$.)
\end{remark}

For the proof of the claim on towers, assume equivalent forward deterministic $\omega$-automata $\Aut A$ and $\Aut A'$ are given and consider any tower $\langle C_0, \dots, C_{m-1}\rangle$ in $\Aut A$, say a positive one; the argument is symmetric for a negative one. Let $q$ be a state in $C_{m-1}$ and, for every $i < m$, let the word $v_i$ be a label for a loop at $q$ with loop set $C_i$. Further, let $u$ be a word such that $\delta(q_I, u) = q$. Then $\delta(q_I, uw) = q$ for every word $w \in (v_0 + \dots + v_{m-1})^*$. Moreover, whether a word $u v_{i_0} v_{i_1} \dots$ is accepted is determined by the least index occurring infinitely often among $i_0, i_1, \dots$. Let $k$ be greater than the number of states of $\Aut A'$ and consider the words $w_i$ defined inductively by $w_0 = \epsilon$ and $w_{i+1} = (v_{m-i-1} w_i)^k$, for $i < m$. Then, using Remark~\ref{thm:loops}, we find that for each $i < m$, some power of $w_{i+1}$ is the label of a loop at $\delta^*(q_I', u w_m)$ and has the same sign as $C_{m-i-1}$. So the reverse sequence of the loop sets forms a positive tower in~$\Aut A'$ of height~$m$.\qed

There is a strong relationship between towers and walls on one side and topological aspects of $\omega$-languages on the other side, see, for instance, \cite{perrin-pin-infinite-words-2004}.

\subsection{The parity index}
\label{sec:parity-index}

From Theorem~\ref{thm:wagner} it follows that, in particular, the greatest height of a tower in a forward deterministic automaton is characteristic for the language recognized. This number is intimately connected with the number of priorities needed by a forward deterministic parity automaton to recognize the same language. To make this more precise we say a parity automaton uses $n$ priorities if $n$ is the maximum of the number of priorities occurring in any strongly connected component of the automaton. Given a regular $\omega$-language the smallest number of priorities used in any forward deterministic parity automata recognizing the language is its parity index.\index{parity index!of an $\omega$-language} 

\begin{corollary}[parity index]
 The greatest height of a tower in a given forward deterministic $\omega$-automa\-ton is exactly the parity index of the language recognized.
\end{corollary}

That the parity index is at least the greatest height of a tower follows from Theorem~\ref{thm:wagner}. For the converse, reconsider the construction from Section~\ref{sec:lar} that turns a Muller automaton into an equivalent parity automaton. Essentially, the Muller automaton without recurrence condition is cascaded with the LAA (latest appearance automaton) and augmented by a parity condition. It is enough to adjust the latter as follows. A state $\langle q, v\$v'\rangle$ is assigned the value $n - l + o$ where 
\begin{inparaenum}[(i)]
\item the number $l$ is the greatest height of a tower ending in a loop with loop
 set $\Occ{v'}$ and 
\item the number $o < 2$ is chosen in a way such that $n - l + o$ is even if the loop is positive and else odd.\qed 
\end{inparaenum}

The Rabin index of a regular $\omega$-language \cite{wagner-1977} is a similar but somewhat coarser measure.\index{Rabin index!of an $\omega$-language}

\subsection{Forward deterministic weak automata}

In terms of the above complexity measure---parity index---the simplest forward deterministic automata that can be considered are the ones with parity index~$1$; these automata are exactly the forward deterministic weak automata. 

On one hand, weak automata are indeed weak in the sense that the class of languages recognized by them is small, for instance, $(01)^\omega$ cannot even be recognized by such automata. In fact, there is a simple characterization of languages recognized by forward deterministic weak automata.

\begin{remark}\cite{staiger-wagner-automatentheoretische-automatenfreie-charakterisierungen-1974}
 \label{thm:weak-characterization}
 An $\omega$-language can be recognized by a forward deterministic weak automaton over some alphabet $A$ if, and only if, it is a boolean combination of languages of the form $UA^\omega$ where $U$ is a regular language of finite words.
\end{remark}

On the other hand, weak automata have some properties which general $\omega$-automata are lacking. One interesting property is described in Theorem~\ref{thm:initial-congruence}. Another property has to do with their determinization:

\index{conditional determinization!of an $\omega$-automaton}
\begin{theorem}[conditional determinization \cite{boigelot-jodogne-wolper-effective-decision-procedure-linear-arithmetic-over-integers-reals-2005}]
 \label{thm:forward-deterministic-weak} 
 If an $\omega$-language is recognized by some forward deterministic weak automaton, then a variant of the breakpoint construction can be used to transform a forward nondeterministic weak automaton recognizing the language into an equivalent forward deterministic weak automaton.
\end{theorem}

\subsection{Loops in backward deterministic automata}
\label{sec:backward-loops}

The requirement that in a backward deterministic automaton there is exactly one recurring run for every $\omega$-word over the given alphabet is a very strong one, which has interesting implications.

\begin{proposition}
 \cite{carton-michel-unambiguous-buechi-automata-2003}
 An $\omega$-automaton is backward deterministic if, and only if, its transition relation is backward deterministic and every nonempty finite word is the label of a positive loop at exactly one state.
\end{proposition}

For the proof, assume a backward deterministic automaton is given and let $u$ be a nonempty finite word. If it is the label of a positive loop at two distinct states $q$ and $q'$, then there are at least two recurring runs of the automaton on $u^\omega$---a contradiction. So $u$ can only be the label of a positive loop at at most one state. Since there is a recurring run of the automaton on $u^\omega$, there is some $k$ such that $u^k$ is the label of a loop at state $q$. If $\delta(u,q) \neq q$, then $u^k$ would also be the label of a loop at $\delta(u,q)$ and there would be two recurring runs for $u^\omega$---a contradiction. So $u$ is the label of a positive loop at at least one state. 

For the converse, assume every nonempty finite word is the label of a positive loop at exactly one state. Then every periodic word over the given alphabet has a recurring run and hence every ultimately periodic word over the same alphabet has so, too. In other words, the set of all words without recurring run is a regular $\omega$-language without ultimately periodic words. From Remark~\ref{thm:basic}(1), we can conclude this set is empty. This shows that for every $\omega$-word there is at least one recurring run. By way of contradiction, assume there are two distinct recurring runs on a given $\omega$-word~$u$, say $r$ and $r'$. Because of the backward deterministic transition relation there must be some $i$ such that $r(j) \neq r'(j)$ for all $j \geq i$. 
As a consequence, there are positions $i$ and $j$ such that 
\begin{inparaenum}[(i)]
\item $i<j$, $r(i) = r(j)$, $r'(i) = r'(j)$, and $r(i) \neq r'(i)$, and 
\item $r(i)u(i)r(i+1) \dots r(j)$ as well as $r'(i)u(i)r'(i+1) \dots r'(j)$ are positive loops at different states with the same label.
\end{inparaenum}
This is a contradiction to the assumption.\qed

The above proposition, in combination with the final syntactic congruence (defined subsequent to Theorem~\ref{thm:initial-congruence}), can be used to classify regular $\omega$-languages using backward deterministic $\omega$-automata, see \cite{preugschat-wilke-effective-characterizations-of-simple-fragments-of-temporal-logic-using-carton-michel0automata-2013}.

\section{Alternation}

$\omega$-Automata are often used in the context of two-player games of infinite duration played on graphs, and results on such games are useful tools for obtaining results on $\omega$-automata. For infinite trees, alternation is an even more important concept.

\subsection{Games of infinite duration with regular winning conditions}
\label{sec:games}

In this section, the fundamentals of games of infinite duration with regular winning conditions are recalled. Remark~\ref{thm:regular-winning} is one of the prime applications of \emph{forward deterministic} $\omega$-automata; in combination with Theorem~\ref{thm:positional}, it explains why the parity condition is so important.

The players of a two-player game of infinite duration\index{game!two-player!of infinite duration} played on graphs are called Zero and One; a game is given by a set $V$ of vertices, a set $E \subseteq V \times V$ of edges, a set $V_0$ of vertices owned by Zero, and a winning condition $W \subseteq V^\omega$. A play of such a game starting in some vertex $v_I$ is a maximal path through the graph starting with the vertex $v_I$; the idea is that a pebble is moved over the edges of the graph from one vertex to the next, starting with the pebble on vertex $v_I$, and Zero moving in her vertices and One moving in the vertices owned by him, which are the ones in $V \setminus V_0$. A play\index{play!of a two-player game of infinite duration} is winning for Zero if the path is either finite and its last vertex belongs to One (that is, One cannot move anymore) or infinite and belongs to $W$; else it is winning for One.

When a player has a strategy for winning the plays starting in a particular vertex, the player is said to win the game starting in this vertex.\index{win!of a two-player game of infinite duration} The set of such vertices is called his or her winning region.

Often, a winning condition is a regular $\omega$-language. More precisely, a coloring function $c \colon V \to C$ into a finite set of so-called colors and a regular $\omega$-language~$L$ over~$C$ are given; the winning condition $W$ is determined by $W = \{u \in V^\omega \mid c \circ u \in L\}$. One speaks of a regular winning condition.\index{regular winning condition!of a two-player game of infinite duration}

A consequence of Martin's theorem \cite{martin-borel-determinacy-1975} and McNaughton's result \cite{mcnaughton-testing-and-generating-infinite-sequences-1966} is:

\index{determinacy!regular}\index{regular determinacy}
\begin{theorem}[regular determinacy]
 \label{thm:determinacy}
 Given a game with a regular winning condition and a vertex in this game, either Zero or One wins the game starting in this vertex. The game is said to be determined in the vertex.
\end{theorem}

A game is called a parity game if there is a function $\pi \colon V \to [n]$ such that $u \in W$ if, and only if, $\liminf_i \pi(u(i))$ is even. Hence, parity games can be viewed as games with a regular winning condition. From the fact that every regular $\omega$-language is recognized by some forward deterministic parity automaton, one can derive:

\index{embedding!of a game into a parity game}
\begin{remark}
 \label{thm:regular-winning}
 Every game with a regular winning condition can be embedded into a game with a parity winning condition.
\end{remark}

To understand what exactly this means assume a game with a regular winning condition as described above and a forward deterministic parity automaton $\Aut A$ recognizing the language $L$ are given. Consider the modified game with vertex set $V \times Q$, Zero's vertex set $V_0 \times Q$, edge set $\{\langle \langle v,q\rangle, \langle v', \delta(q, c(v'))\rangle \rangle \mid \langle v, v'\rangle \in V \text{ and } q \in Q\}$, and priority function $\langle v, q\rangle \mapsto \pi(q)$. Because the automaton $\mathscr A$ is forward deterministic, playing in the original game starting from a vertex $v$ is exactly the same as playing in the new game starting from the vertex $\langle v, q_I\rangle$. In particular, Zero wins the former game in a vertex $v$ if, and only if, she wins the latter game in $\langle v, q_I\rangle$. In other words, without loss of generality, only games with parity winning conditions need to be considered when regular winning conditions are used.

In general, regular winning conditions may require a player to remember a certain amount of information in order to win. For instance, if the winning condition demands that Zero visits the vertices $v_1$ and $v_2$ infinitely often in the graph 
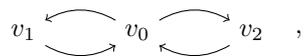
\begin{center}
 \begin{tikzpicture}[baseline=(b.center)]
	 \node (a) {$v_0$};
	 \node[left of=a,node distance = 15mm] (b) {$v_1$};
	 \node[right of=a,node distance = 15mm] (c) {$v_2$};
	 \draw[->] (a) to [bend right] (b);
	 \draw[->] (b) to [bend right] (a);
	 \draw[->] (c) to [bend left] (a);
	 \draw[->] (a) to [bend left] (c);
 \end{tikzpicture} \enspace,
\end{center}
where $v_0$ is her vertex, then she cannot base her decision what to do in vertex $v_0$ only on the fact that she is in that vertex. (Of course, when she remembers where she moved previously, she can alternate and win.) Opposed to this, if the winning condition demands that Zero visits $v_2$ infinitely often, she only needs to follow the rule ``if in vertex $v_0$, go to vertex~$v_2$''---her decision what to do next is only based on the current vertex.

A uniform positional winning strategy for Zero is a function $(W_0 \cap V_0) \to W_0$, where $W_0$ is Zero's winning region, such that no matter where in $W_0$ a play starts, if Zero moves as determined by the function, then the resulting play is winning for Zero. For One, the definition is symmetric.

\index{positional strategy!in a two-player game of infinite duration}
\begin{theorem}[positional strategies \cite{emerson-jutla-tree-automata-mu-calculus-determinacy-1991}]
 \label{thm:positional}
 In every parity game, both players have a uniform positional winning strategy.
\end{theorem}

\subsection{State- and transition-controlled alternating automata}
\label{sec:alternation}

In general, an alternating automaton is an automaton where acceptance depends on the full computation tree on a given word, more precisely, such an automaton provides means for specifying that a given word is accepted if, and only if, a certain subgraph of the full computation tree exists. At one extreme, when this subgraph is required to be a rooted path, then the automaton is nothing else than a conventional automaton.

\index{alternating $\omega$-automaton}\index{$\omega$-automaton!alternating}
For $\omega$-automata, essentially two variants of alternating automata have been studied: in one variant, alternation is specified by partitioning the state space \cite{miyano-hayashi-alternating-finite-automata-on-omega-words-2-1984}; in the other variant, alternation is specified by complex transition formulas \cite{muller-schupp-alternating-automata-infinite-trees-1987}.

\index{state-controlled alternating $\omega$-automaton}\index{$\omega$-automaton!alternating!state-controlled}
In the state-controlled variant the state space $Q$ is partitioned into a set $E$ of existential states and a set $U$ of universal states (where either set could be empty) and the set of initial states is either a subset of $E$ or of $U$. A run of the automaton on a word $u$ is a prefix-closed set $T \subseteq Q^*$, which should be thought of as a tree satisfying the following properties for every vertex $vq \in T$:
\begin{asparaitem}
\item if $q \in E$, then there exists a state $q'$ such that $vqq' \in T$ and $\langle q, u(|v|), q'\rangle \in \Delta$; 
\item if $q \in U$, then $vqq' \in T$ for every $q'$ such that $\langle q, u(|v|), q'\rangle \in \Delta$.
\end{asparaitem}
The run is initial if either $Q_I \subseteq E$ and $Q_I \cap T \neq \emptyset$ or $Q_I \subseteq U$ and $Q_I \subseteq T$; it is recurring if every word $r \in Q^\omega$ whose finite prefixes all belong to $T$ (every infinite rooted path through~$T$) is recurring in the sense of the given transition condition. This means, in particular, if $E = Q$ and the set of initial states is existential, then the automaton can be viewed as an ordinary $\omega$-automaton. It is said to be a universal automaton if $U = Q$.\index{universal $\omega$-automaton}\index{$\omega$-automaton!universal}

From the closure under complementation of the class of $\omega$-languages recognized by B\"{u}chi automata, one obtains immediately:

\begin{remark}
 \label{thm:universal-co-buechi}
 Every regular $\omega$-language is recognized by a universal co-B\"{u}chi automaton.
\end{remark}

\index{transition-controlled alternating $\omega$-automaton}\index{$\omega$-automaton!alternating!transition-controlled}
In the transition-controlled variant, the transition relation is replaced by a transition function $\delta \colon Q \times A \to M(Q)$, where $M(Q)$ is the set of all expressions built from states, the connectives $\vee$ (``or'') and $\wedge$ (``and''), and the boolean constants $0$ (``false'') and $1$ (``true''). For instance, $q \wedge (q' \vee q'')$ could be a value of the transition function. The set of initial states is replaced by an expression from $M(Q)$. Again, a run is a prefix-closed set \mbox{$T \subseteq Q^*$}, but this time satisfying the following condition. For each vertex $vq \in T$, the set $\{q' \mid vqq' \in T\}$ satisfies the expression $\delta(q, u(|v|))$, where satisfaction is defined in the obvious way. A run is initial if $\{q \in Q \mid q \in T\}$ satisfies the initial condition. Being recurring is defined as above.

\begin{remark}
 A state-controlled alternating $\omega$-automaton can be viewed as a transition-controlled alternating $\omega$-automaton.
\end{remark}

More precisely, for every existential state $q$ one sets $\delta(q, a) = \bigvee \{q' \mid \langle q, a, q'\rangle \in \Delta\}$, and for every universal state $q$ one sets $\delta(q, a) = \bigwedge \{q' \mid \langle q, a, q' \rangle \in \Delta\}$; the set of initial states is converted into an initial condition in the same way; the recurrence condition does not need to be changed.

\subsection{Alternating automata and games}

Given a state-controlled alternating automaton as above and an $\omega$-word $u$ over the same alphabet, the question whether $u$ is accepted by the automaton can be viewed as the question whether Zero wins a certain game, the so-called automaton game for $u$. The vertices of this game are pairs of the form $\langle q, u'\rangle$, where $u'$ is a suffix of $u$; such a vertex belongs to Zero if, and only if, $q$ is existential; there is an edge from $\langle q,u'\rangle$ to $\langle q', u''\rangle$ if $\langle q,u'(0), q'\rangle \in \Delta$ and $u'' = u'(1)u'(2) \dots$; the recurrence condition is adapted in the straightforward fashion, based on the state in the first component. 

\begin{remark}
 \label{thm:alternation-game}
 A state-controlled alternating $\omega$-automaton with existential [universal] initial states accepts a word $u$ if, and only if, Zero wins the automaton game for $u$ in some [every] vertex in $Q_I \times \{u\}$.
\end{remark}

From Theorem~\ref{thm:determinacy}, which states that the games that occur in this fashion are determined, one can derive that a word is not accepted if, and only if, One has a winning strategy. This is equivalent to saying that the dual automaton accepts the word, where dualizing an automaton has the obvious meaning:\index{dual!of an alternating $\omega$-automaton}\index{$\omega$-automaton!alternating!dual of} existential and universal states exchange their roles and the recurrence condition is replaced by its negation. In other words, complementation is a trivial problem for alternating automata.

\begin{proposition}[complementing alternating automata]
 \label{thm:complementation-alternation}
 The dual of a state-controlled alternating $\omega$-automaton recognizes the complement of the language recognized by the given automaton.
\end{proposition}

Remark~\ref{thm:alternation-game} and Proposition~\ref{thm:complementation-alternation} hold true for transition-controlled alternating $\omega$-au\-tom\-a\-ta as well, but the definition of the automaton game and the dualization process need to be adapted. The vertices of the automaton games are of the form $\langle \phi, u'\rangle$ where $\phi$ is a subformula of some value of the transition function. In the dualization process, the values of the transition function and the initial condition are dualized.

\subsection{From alternating automata to nondeterministic ones}

Alternating $\omega$-automata can be exponentially more concise than ordinary ones, just as in the finite-word setting~\cite{drusinsky-harel-on-the-power-of-bounded-concurrency-i-finite-automata-1994}, but with regard to expressive power there is no difference. This is a major application of complementing $\omega$-automata.

\begin{theorem}[from alternating to nondeterministic \cite{miyano-hayashi-alternating-finite-automata-on-omega-words-2-1984}]
 For every alternating $\omega$-au\-tom\-a\-ton there exists an equivalent nondeter\-min\-istic B\"{u}chi automaton.
\end{theorem}

To prove this, first observe that it is enough to consider alternating parity automata, because any Muller condition can be turned into a parity condition as described in the proof of Theorem~\ref{thm:muller-to-parity}. 

By Theorem~\ref{thm:positional}, parity games have uniform positional winning strategies. It follows that if there is an accepting run (recall that runs are trees) of an alternating parity automaton on a given word, then there is also an accepting subgraph of the run DAG, where this is defined in the obvious way. Checking that in a subgraph of a run DAG \emph{all} rooted paths are recurring can be done using an appropriate $\omega$-automaton, as explained in what follows. 

Consider the nondeterministic parity automaton over the alphabet $\wp(Q \times Q)$, with state set $Q$, initial set $Q_I$, transition relation $\{\langle q, a, q'\rangle \mid \langle q,q'\rangle \in a\}$, and parity condition $\pi + 1$. This automaton accepts a word $u$ if the DAG which is obtained by collating the letters of $u$ contains \emph{some} initial rooted path starting in an initial state and not satisfying the parity condition of the original automaton. Any $\omega$-automaton recognizing the complement of the language recognized by this automaton is one that can check the DAG's. 

To sum up, cascading 
\begin{inparaenum}[(i)]
\item an automaton producing a subgraph of a run DAG of a given $\omega$-word satisfying the transition relation and 
\item the above automaton
\end{inparaenum}
yields the desired automaton.\qed

\subsection{Weak alternating automata}
\label{sec:weak-alternation}

\index{weak alternating $\omega$-automaton}\index{$\omega$-automaton!weak alternating}
Remark~\ref{thm:weak-characterization} states that weak deterministic $\omega$-automata only recognize fairly simple $\omega$-languages. This is different for alternating automata:

\begin{theorem}[from alternating to weak alternating]
 \cite{kupferman-vardi-weak-alternating-automata-are-not-that-weak-2001}
 For every alternating B\"{u}chi automaton with $n$ states there exists an equivalent weak alternating automaton with $2n^2$ states.
\end{theorem}

By dualization, it is enough to consider alternating co-B\"{u}chi automata. Theorem~\ref{thm:positional} says that runs of alternating co-B\"{u}chi (and B\"{u}chi) automata can be thought of as run DAG's. The use of rank functions from Section~\ref{sec:dags} leads to the following characterization of when an $\omega$-word $u$ is accepted by a transition-controlled co-B\"{u}chi alternating automaton with $n$ states. There exists a tree $T \subseteq (Q \times [2n])^*$ satisfying the following conditions.
\begin{inparaenum}[(i)]
\item The set $\{q \in Q \mid \langle q,c \rangle \in T \text{ for some $c< 2n$}\}$ satisfies the initial condition.
\item Whenever $v \langle q, c \rangle \in T$, then $\{q' \in Q \mid v \langle q, c \rangle \langle q', c'\rangle \in T \text{ for some $c' < 2n$}\}$ satisfies $\delta(q,u(|v|))$. 
\item There is no vertex $v \langle  q, 2j+1 \rangle \in T$ with $q \in B$. 
\item When $v \langle q, c \rangle \langle q', c'\rangle \in T$, then $c \geq c'$.
\item For every rooted path $\langle q_0, c_0 \rangle \langle q_1, c_1 \rangle \dots$ there exists some $i$ such that $c_i, c_{i+1}, \dots$ are all odd. 
\end{inparaenum}
This can be used to construct a transition-controlled weak alternating automaton with state set $Q \times [2n]$; the initial condition and the transition function are adapted from the given automaton in a straightforward fashion; the B\"{u}chi set consists of all states with an odd second component.\qed

It should be noted that the breakpoint construction can be used to convert a weak alternating B\"{u}chi automaton into an equivalent nondeterministic one.

\subsection{Simulation relations and simulation games}

One way to compare automata with each other, more precisely, to compare their internal structure, is to use simulation relations, or, more generally, simulation games.\index{simulation relation!for $\omega$-automata}\index{simulation game!for $\omega$-automata}

A simple approach is to say that a B\"{u}chi automaton $\mathscr A'$ forwardly simulates a B\"{u}chi automaton $\mathscr A$\index{forward simulation!of $\omega$-automata} if there is a relation $\sigma \subseteq Q \times Q'$ such that the following three conditions are satisfied.
\begin{inparaenum}[(i)]
\item For every $q \in Q_I$ there is some $q' \in Q_I'$ such that $\langle q , q' \rangle \in \sigma$.
\item For all $\langle q, q'\rangle \in \sigma$ and $\langle q, a, r\rangle \in \Delta$ there is some $r' \in Q'$ such that $\langle r, r'\rangle \in \sigma$ and $\langle q', a, r'\rangle \in \Delta'$.
\item For all $\langle q, q'\rangle \in \sigma$, if $q \in B$, then $q' \in B'$.
\end{inparaenum}

The important observations concerning this definition are:\index{direct simulation!of $\omega$-automata}
\begin{theorem}[direct simulation 
 \cite{dill-hu-wong-toi-checking-for-language-inclusion-using-simulation-preorder-1991}]
 \label{thm:simulation}
 \begin{compactenum}
 \item If a B\"{u}chi automaton $\mathscr A'$ simulates a B\"{u}chi automaton $\mathscr A$, then the language recognized by $\mathscr A$ is a subset of the language recognized by $\mathscr A'$.
 \item Whether a B\"{u}chi automaton $\mathscr A'$ simulates a B\"{u}chi automaton $\mathscr A$ can be determined in time linear in the product of the sizes of $\mathscr A$ and $\mathscr A'$.
 \end{compactenum}
\end{theorem}

As a consequence, simulation relations can be used for efficient (but incomplete) inclusion tests.

The requirement that a B\"{u}chi state in the simulating automaton match a B\"{u}chi state in the simulated automaton right away is very strong. For inclusion to hold, it would be enough if a B\"{u}chi state in the simulated automaton is matched by a B\"{u}chi state in the simulating automaton at a later position. This is captured by the notion of delayed simulation,\index{delayed simulation!of $\omega$-automata} which is best phrased in terms of a certain two-player game, where one of the players is called Duplicator and tries to show that simulation is given, whereas the other is called Spoiler and tries to show that this is not the case. 

More precisely, the game determines whether a state in a B\"{u}chi automaton $\mathscr A'$ delayed simulates a state in a B\"{u}chi automaton $\mathscr A$. When a play of the game starts, there is a pebble on each of the two states in question. In every round of the game, first Spoiler is required to move the pebble on $\mathscr A$ over some transition and then Duplicator is required to move the other pebble (the pebble  in $\mathscr A'$) over some transition with the same label. If one of the players cannot move anymore, this player looses early. If an infinite play emerges, then Duplicator wins if, and only if, the following holds: whenever Spoiler visits a B\"{u}chi state in some round, Duplicator visits a B\"{u}chi state in the same or in a later round. The state in $\Aut A'$ delayed simulates the state in $\Aut A$ if Duplicator has a winning strategy in the game just described. The automaton $\mathscr A'$ delayed simulates the automaton $\mathscr A$ if every initial state of $\mathscr A$ is simulated by some initial state of $\mathscr A'$. Observe that the above game can be viewed as a game of infinite duration with a regular winning condition as described in Section~\ref{sec:games}.

Theorem~\ref{thm:simulation} carries over to delayed simulation, only the complexity of computing delayed simulation is higher \cite{etessami-schuller-wilke-fair-simulation-relations-parity-games-and-state-space-reduction-for-buechi-automata-2005}.

For purposes of state-space reduction, it useful to study simulation in both directions: if one state [delayed] simulates another one and vice versa, the states are said to mutually [delayed] simulate each other. These relations are, indeed, equivalence relations and have a useful property:

\begin{theorem}[quotienting   \cite{etessami-schuller-wilke-fair-simulation-relations-parity-games-and-state-space-reduction-for-buechi-automata-2005}]
If, in a quotient of a B\"{u}chi automaton with regard to the mutual [delayed] simulation relation, initial and B\"{u}chi states are chosen appropriately, then the resulting automaton is equivalent to the given one.
\end{theorem}

This gives, in effect, two polynomial-time algorithms for reducing the state space of B\"{u}chi automata, one less efficient than the other, but producing smaller automata. Finding and even approximating minimum-size B\"{u}chi automata is PSPACE-hard, in fact, this is independent of the type of the automaton, because results from finite-state automata on finite words ~\cite{gramlich-schnitger-minimizing-nfas-and-regular-expressions-2007} carry over in a straightforward fashion.

In principle, one could also work with bisimulation rather than mutual simulation, but this gives, in general, worse reductions. 

Much effort has gone into finding coarser relations for state-space reductions, and there are various ways of approaching this: letting Duplicator match with more than just one pebble, relaxing the winning condition for Duplicator further, considering backward simulation, and so on.

\section{Applications in logic}

$\omega$-Automata were introduced in the late fifties in the context of mathematical logic, more precisely, B\"{u}chi automata first showed up in \cite{buechi-decision-method-restricted-second-order-arithmetic-1962} (in disguise) and were used there as a tool for proving that theories of specific structures are decidable. From a modern point of view, B\"{u}chi showed that the structures are $\omega$-automatic~\cite{hodgson-decidabilite-par-automata-fini-1983}.

\subsection{\texorpdfstring{$\omega$}{ω}-Automatic structures}

Assume a first-order structure $\mathfrak S$ consisting of a universe $U$ and a family $\{R_i\}_{i \in I}$ of relations, say $R_i$ having arity $n_i$, are given; the question is whether the theory of this structure is decidable. A good example are the real numbers with the ternary relation ``addition'', the predicate ``is positive'', and the predicate ``is power of 2''. 

An $\omega$-automatic presentation\index{$\omega$-automatic}\index{structure!$\omega$-automatic}\index{presentation!$\omega$-automatic}\index{$\omega$-automatic presentation} of a structure $\mathfrak S$ as above is given by an alphabet $A$, an $\omega$-automaton $\mathscr U$ over $A$, an $\omega$-automaton $\mathscr E$ over $A \times A$, and, for each $i \in I$, an $\omega$-automaton $\mathscr R_i$ over $\bigtimes_{i<n_i} A$. It is required that there exists an onto function $f \colon \text L(\mathscr U) \to U$ such that the following conditions are satisfied:
\begin{asparaitem}
\item For all $u, v \in L(\mathscr U)$, $f(u) = f(v)$ if, and only if, $u * v \in L(\mathscr E)$.
\item For all $i \in I$ and $u_0, \dots, u_{n_i - 1} \in L(\mathscr U)$, $u_0 * \dots * u_{n_i-1} \in L(\mathscr R_i)$ if, and only if, $\langle f(u_0), \dots, \linebreak f(u_{n_i-1})\rangle \in R_i$.
\end{asparaitem}
To extend the above example, one can start with an $\omega$-automaton $\mathscr U$ that accepts exactly the $\omega$-words representing real numbers as described in Section~\ref{sec:omega-words}. Then $\mathscr E$ must be constructed in a way such that it identifies representations of identical numbers. Finally,  $\omega$-automata representing the three respective relations must be found. A simple automaton is the ``is power of $2$'' automaton, which only checks that there is exactly one occurrence of~$1$ and that this occurrence is not in position~$0$ (because otherwise the number represented would be $0$, more precisely, $-0$, which is not a power of~$2$).

The fundamental result about $\omega$-automatic structures is:

\index{first-order theory!of $\omega$-automatic structures}
\begin{theorem}[$\omega$-automatic structures
 \cite{hodgson-decidabilite-par-automata-fini-1983}]
 \label{thm:automatic}
 The first-order theory of every $\omega$-automatic structure is decidable.
\end{theorem}

The reason for this is that, by induction, one can show that for every first-order formula in the respective vocabulary one can construct an $\omega$-automaton that recognizes the representations of the satisfying assignments. When $\phi = \phi(x_0, \dots, x_{n-1})$ is a formula with all of its free variables among $x_0, \dots, x_{n-1}$, then a word of the form $u_0 * \dots * u_{n-1}$ represents a satisfying assignment if $u_i \in L(\mathscr U)$ for every $i<n$ and $\mathfrak S, f(u_0), \dots, f(u_{n-1}) \models \phi(x_0, \dots, x_{n-1})$.

For the base case, there is almost nothing to show, because this is part of the definition of $\omega$-automatic structure. For the induction itself it should be noted that disjunction can essentially be viewed as union, negation as complementation, and existential quantification as projection. All these operations can easily be implemented effectively on $\omega$-automata. In other words, there is an effective procedure that, given a closed formula, constructs an $\omega$-automaton over the unary alphabet, $\bigtimes_{i<0} A$, which accepts some word, more precisely, the word $\langle \rangle^\omega$, if, and only if, the formula is true in the given structure. Nonemptiness can be verified effectively for $\omega$-automata, see Remark~\ref{thm:basic}(2).\qed

An important example for this theorem, already mentioned in B\"{u}chi's seminal paper, is the one described above: the real numbers with addition and the ``is positive'',  ``is power of two'', and ``is an integer'' predicates.

Another example from B\"{u}chi's original work is the monadic second-order theory of the natural numbers with successor, more precisely: the structure is the set of natural numbers endowed with the successor predicate; in the vocabulary of the logical language there are, in addition to what is part of a suitable first-order language (symbols for disjunction, negation, existential quantification, the binary successor relation, variables for natural numbers),  variables for sets of natural numbers, a symbol for ``is element of'', and a symbol for existential quantification of set variables. At first glance, this does not look like a situation where Theorem~\ref{thm:automatic} can be applied, but it actually can: a formula in the above logic can be translated in a straightforward fashion into a first-order formula for the structure with the power set of the natural numbers as universe and endowed with the ``is singleton'' predicate, the binary relation ``is subset of'', and the binary relation ``every element of \dots\ has a successor in \dots''.

\index{S1S}\index{monadic second-order theory!of one successor}
\begin{theorem}[decidability of S1S
 \cite{buechi-decision-method-restricted-second-order-arithmetic-1962}]
 The monadic second-order theory of the natural numbers with the successor predicate is decidable. 
\end{theorem}

From the point of view of $\omega$-automata theory, there are several applications in logic which are of particular interest. Two of them are discussed in what follows.

\subsection{Temporal logic}

Temporal logic\index{temporal logic} comes in many different flavors. The version that is most often considered and also most amenable to being dealt with using $\omega$-automata is the one where the temporal operators used are future operators (next, X; eventually in the future, F; always in the future, G; until, U; release, R) and the time domain is discrete, more precisely, where the time domain is $\omega$, the set of natural numbers. In such a context, a temporal variable, here denoted $p_i$, is assigned a set of natural numbers, the points in time where the variable is true. So if the variables occurring in a given formula $\phi$ are among $p_0, \dots, p_{n-1}$, then the models of this formula can be viewed as $\omega$-words over the alphabet $\wp(\{p_0, \dots, p_{n-1}\})$. 

For instance, the set of models of the formula $\text G(p_0 \rightarrow \text F p_1)$, which is read  ``now and always in the future, if $p_0$, then $p_1$ at the same time or some point later'', can be viewed as the set of $\omega$-words over $\wp(\{p_0, p_1\})$ with the property that whenever $p_0$ is an element of a letter at some position, then $p_1$ belongs to the same letter or some other letter in a position to the right. 

The starting point for constructing an $\omega$-automaton recognizing the set of models of a given formula  is the observation that whether a formula of temporal logic is true in some point in time only depends on 
\begin{inparaenum}[(i)]
\item which of its strict subformulas hold true in this and the next point in time and
\item whether the formula itself holds true in the next point in time.
\end{inparaenum}
So a suitable automaton can guess, for each point in time, which subformulas are true and then verify its guessing locally in a backward deterministic fashion. For some temporal operators, it is important though to also verify certain conditions globally. For instance, it is true that the formula $\text F \phi$ holds true in position~$i$ if, and only if, $\phi$ holds true in position~$i$ or $\text F \phi$ holds true in position $i+1$, and $\text F \phi$ holds true in position $i+1$ if, and only if, $\phi$ holds true in position $i+1$ or $\text F \phi$ holds true in position $i+2$, and so on, but, clearly, the formula $\phi$ must become true at some point. Such a global condition can be captured by an appropriate recurrence condition. The initial states are the ones where the automaton guesses the entire formula to be true. 

The general theorem is as follows:

\begin{theorem}[from temporal logic to automata
\cite{vardi-wolper-sistla-reasoning-infinite-computation-paths-1983,vardi-wolper-reasoning-infinite-computation-paths-1994}]
 Every temporal formula with~$n$ subformulas can be translated into an equivalent backward deterministic generalized B\"{u}chi automaton with at most $2^n$ states and as many B\"{u}chi sets as there are subformulas with leading temporal operator $\rm F$ or $\rm U$.
\end{theorem}

This implies, in particular, that satisfiability and validity of temporal formulas as well as model checking temporal formulas over finite-state system with fairness conditions are problems in PSPACE \cite{sistla-clarke-complexity-propositional-linear-temporal-logics-1985}. 

Future linear-time temporal formulas can be translated directly into weak alternating automata (see Section~\ref{sec:weak-alternation}) of a very specific structure; the resulting number of states is the number of subformulas.

\subsection{The additive theory of the reals}
\index{additive theory!of the reals}

Regarding the aforementioned example of the real numbers one can show (by other means than automata-theoretic ones) that the relations definable by formulas in the underlying first-order language are all recognizable by forward deterministic weak automata as introduced in Section~\ref{sec:wagner} (when real numbers are represented as described in Section~\ref{sec:omega-words}). In view of Theorem~\ref{thm:forward-deterministic-weak}, this means that a recursive procedure for constructing automata representing such relations can take advantage of conditional determinization, which is much less complicated than ordinary determinization, and of simple and fast minimization procedures as pointed out in Theorem~\ref{thm:initial-congruence}. This, in the end, leads to feasible decision procedures \cite{boigelot-jodogne-wolper-effective-decision-procedure-linear-arithmetic-over-integers-reals-2005}.

\section{More complex recurrence conditions}

Much effort can and has been put into extending $\omega$-automata like the ones dealt with in this paper, that is, the ones with a finite state space and a recurrence condition based on the states occurring infinitely often in a run. There are finite-state automata working on other infinite objects: other ordinals, the integers, linear orderings in general, and, most notably, infinite trees; there are $\omega$-automata using more complex storage, for instance, $\omega$-automata with stacks; there are probabilistic $\omega$-automata, that is, $\omega$-automata where transitions are taken with certain probabilities; there are timed $\omega$-automata, using clocks and processing infinite sequences of events having a duration; and so on.

Beside all this, Boja\'nczyk and Colcombet suggest in \cite{bojanczyk-colcombet-bounds-in-omega-regularity-2006}  to strengthen the models discussed in this chapter by a more powerful mechanism for defining recurrence, allowing a finer analysis of what happens ``in the infinite''. In their model, every automaton has a finite number of counters. A transition is of the form $\langle q, a, \alpha, q'\rangle$ where $\alpha$ is a function assigning to each counter no action or one of the following two:
\begin{inparaitem}[]
\item ``inc''---increment the counter by one;
\item ``prt\&res''---output (print) the counter value and then reset the counter.
\end{inparaitem}
So, for every counter, a finite or infinite sequence of natural numbers, its recurrence sequence, is produced in each run. The recurrence condition is a boolean combination of conditions of the form $\liminf c = \infty$ and $\limsup c = \infty$, with $c$ standing for a counter. A run of such an automaton is recurring if every recurrence sequence is infinite and they all satisfy the recurrence condition.

A good example for a non-regular $\omega$-language which can be recognized by such an automaton is the language $L_\text{ub0}$ (see Section~\ref{sec:saturation}) of all $\omega$-words of the form $0^{i_0}10^{i_1}1\dots$ where $\limsup_{j \to \infty} i_j = \infty$. This language is recognized by an automaton with one counter, say~$c$:\\ 
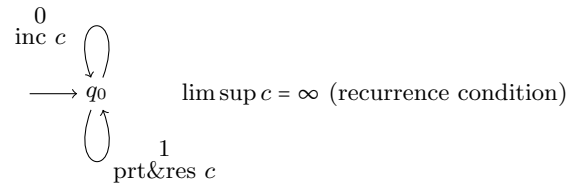
\hspace*{\fill}
 \begin{small}
	 \begin{tikzpicture}[node distance = 20mm]
		 \node[] (s) {}; 
		 \node (s0) [right of=s,node distance=10mm] {$q_0$}; 
		 \draw[->] (s) -- (s0);     
		 \draw[scale=1.5,my loop] (s0) to node[above, left=3mm, align=center] {$0$\\[-1mm] inc $c$} (s0); 
		 \draw[scale=1.5,rotate=180,my loop] (s0) to node[above, right=1mm, align=center] {$1$\\[-1mm] prt\&res $c$} (s0); 
		 \node [right of=s, right] {$\limsup c = \infty$ (recurrence condition)};
	 \end{tikzpicture}
 \end{small}
\hspace*{\fill}

\smallskip\noindent
\textbf{Acknowledgment} \ I am grateful to Christof, Olivier, Sebastian, and my master students for insightful comments, to Wolfgang for his constant support, and to Jean-\'{E}ric for making me write this paper.

\bibliographystyle{abbrv}
\addcontentsline{toc}{section}{References}

\bibliography{abbrevs,ams-perrin-pin,/home/thomas/seabox/bigbib/ti-bib-db}

\end{document}